\documentclass[aps,prd,reprint,twocolumn,groupedaddress]{revtex4}
\usepackage{amsmath,amsfonts,amssymb,bm}
\usepackage{graphicx}
\usepackage{color}
\usepackage{subfigure}
\usepackage{multirow}
\usepackage{textcomp}
\usepackage{slashed}
\usepackage{makecell}
\usepackage{enumitem}

\definecolor{purple}{rgb}{0.5,0,0.5}
\definecolor{blue}{rgb}{0.0,0,0.9}

\usepackage[colorlinks=true, pdfstartview=FitV, linkcolor=purple, citecolor= purple, urlcolor=blue]{hyperref}

\begin{document}


\title{Single-quark electromagnetic form factors of charmonium up to $J=2$}




\author{Jian Huang$^1$, Muyang Chen$^{1,2,3}$\footnote {E-mail: muyang@hunnu.edu.cn}, Xian-Hui Zhong$^{1,2,3}$\footnote {E-mail: zhongxh@hunnu.edu.cn}}

\affiliation{ $^1$Department of Physics, Hunan Normal University, and Key Laboratory of Low-Dimensional Quantum Structures and Quantum Control of Ministry of Education, Changsha 410081, China}

\affiliation{ $^2$Synergetic Innovation Center for Quantum Effects and Applications (SICQEA),
Hunan Normal University, Changsha 410081, China}

\affiliation{ $^3$Hunan Research Center of the Basic Discipline for Quantum Effects and Quantum Technologies, Hunan Normal University, Changsha 410081, China}

\date{\today}

\begin{abstract}
We calculate the single-quark electromagnetic form factors of a broad subset of charmonium, including $\eta_c(1S)$, $\eta_c(2S)$, $\chi_{c0}(1P)$, $\chi_{c0}(2P)$, $J/\psi(1S)$, $J/\psi(2S)$, $\chi_{c1}(1P)$, $\chi_{c1}(2P)$, $h_c(1P)$, $h_c(2P)$, $\chi_{c2}(1P)$ and $\chi_{c2}(2P)$, via a relativized quark model. The reference frame dependence of the results is estimated as the computational error. We compare our results with those of the lattice quantum chromodynamics (LQCD), the Dyson-Schwinger equation (DSE) and the basis light front quantization (BLFQ) approaches where available and we find that most of our results agree with the other results. We also predict the single-quark electromagnetic form factors of $\chi_{c0}(2P)$, $\chi_{c1}(2P)$, $h_c(1P)$, $h_c(2P)$, $\chi_{c2}(1P)$ and $\chi_{c2}(2P)$, where no direct comparisons are available.
\end{abstract}


\maketitle


\renewcommand{\thesection}{\Roman{section}}
\renewcommand{\thesubsection}{\Roman{section}-\Alph{subsection}}

\section{Introduction}\label{sec:introduction}

The charmonium system, comprising a bound state of a charm quark and its antimatter counterpart, has long served as a fundamental laboratory for probing the dynamics of quantum chromodynamics (QCD). As a non-Abelian gauge theory, QCD governs the strong interaction responsible for confining quarks into hadrons. The spectrum of charmonium states—including the \(\eta_c\), \(J/\psi\), \(\chi_{cJ}\), and \(h_c\)—provides critical insights into both perturbative and non-perturbative aspects of QCD. Among the various probes of charmonium structure, electromagnetic interactions offer a particularly clean window due to the well-understood nature of quantum electrodynamics (QED).

Form factors are essential quantities that encode the spatial distribution of the constituent quark and antiquark within a composite particle. For charmonium, there have been extensive researches on the two-photon transition form factors like \(\gamma^* \gamma \to \eta_c\) and the hadronic radiative transition form factors like \(\gamma^* J/\psi \to \eta_c\), for example, see Refs. \cite{Eichten2008,Chen2017,Hoferichter2020,Li2022,Colquhoun2023,Ding2025}. The elastic scattering like \(\gamma^* \eta_c \to \eta_c\) is not an observable, because the contribution from the quark and the antiquark cancel out. This gives rise to the concept of single-quark electromagnetic form factors, which describe the response of an individual quark within the meson to an electromagnetic probe, while the other quark acts as a spectator. In the non-relativistic limit, the magnetic moment of a quark is simply \(\mu_q = e_q/(2m_q)\), but bound-state effects and relativistic corrections can significantly alter this picture. Similarly, the electric quadrupole moment provides direct information about the nonspherical charge distribution arising from the orbital motion of the quark or antiquark.

The single-quark electromagnetic form factors of charmonium have been studied by various theoretical approaches, including lattice QCD \cite{Dudek2006,Chen2011,Li2020,Delaney2024}, quark model \cite{Lakhina2006,Arifi2024}, basis light-front quantization \cite{Adhikari2019}, Dyson-Schwinger equation \cite{Maris2007} and a contact interaction model \cite{HernandezPinto2024}. In this work, we present a systematic study of the single-quark electromagnetic form factors of charmonium states via a relativized quark model. Our results highlight how the internal geometry and dynamics of charmonium, from \(J=0\) to \(J=2\), are reflected in the properties of their constituent quarks. Comparing our results with those from other theoretical approaches, whence available, we find most of our results agrees with those results. No direct comparisons are available for the single-quark electromagnetic form factors of $\chi_{c0}(2P)$, $\chi_{c1}(2P)$, $h_c(1P)$, $h_c(2P)$, $\chi_{c2}(1P)$ and $\chi_{c2}(2P)$, and we give our predictions.

This paper is organized as following. In section \ref{sec:model} we introduce the quark model, the mock meson states and the definition of the electromagnetic form factors. In section \ref{sec:resultsCC} we present and comment our results for the sigle quark electromagnetic form factors of the charmonium, including $\eta_c(1S)$, $\eta_c(2S)$, $\chi_{c0}(1P)$, $\chi_{c0}(2P)$, $J/\psi(1S)$, $J/\psi(2S)$, $\chi_{c1}(1P)$, $\chi_{c1}(2P)$, $h_c(1P)$, $h_c(2P)$, $\chi_{c2}(1P)$ and $\chi_{c2}(2P)$. Summary and conclusion is given in section \ref{sec:conclusion}. There are also two appendixes, \ref{sec:appendixA} formulates the derivation of the non-relativistic limit of the relativized mock meson state, and \ref{sec:appendixB} describes the procedure for calculating the single-quark form factors.

\section{Quark model, Mock Meson State and the electromagnetic form factor}\label{sec:model}

In this work we employ the same interaction model and the mock meson state of Ref. \cite{Ding2025} with unchanged parameters, which yields charmonium mass spectra and two-photon transition form factors of $\eta_{c}$ and $\chi_{c0}$ that show excellent quantitative agreement with experimental data and other theoretical methods. For convenience, we briefly outline the model here.

The masses and wave functions are obtained by solving the radial Schrödinger equation,
\begin{equation}\label{eq:SchrödingerEq}
 (T + V - E)\varphi(\bm{r}) = 0,
\end{equation}
where $T$ is the kinetic energy operator, $V$ is the potential between the quarks, $E$ is the energy of this system and $\varphi(\bm{r})$ is the spacial wavefunction, $\bm{r}$ is the relative coordinate the quark and antiquark.  In this paper, boldface symbols denote three-dimensional vectors (e.g. $\bm{r} = \vec{r}$). The relativistic kinetic energy is
\stepcounter{equation}
\begin{equation}\tag{\theequation a}\label{eq:RelKEnergy}
 T = E_{\bm{k}}+\bar{E}_{\bar{\bm{k}}},
\end{equation}
where $E_{\bm{k}}=\sqrt{m^2+\bm{k}^2}$ and $\bar{E}_{\bar{\bm{k}}} = \sqrt{\bar{m}^2 + \bar{\bm{k}}^2}$, $m$ and $\bar{m}$ are the masses of the quark and antiquark, $\bm{k}$ and $\bar{\bm{k}}$ are the 3-momentum of the quark and antiquark.

The potential could be decomposed into
\begin{equation}\label{eq:interaction}
 V = H^{\text{SI}} + H^{\text{SS}} + H^{\text{T}} + H^{\text{SO}}.
\end{equation}
$H^{\text{SI}}$ is the spin independent part, which is composed of a coulombic potential and a linear potential,
\begin{equation}\label{eq:interactionSI}
 H^{\text{SI}} = -\frac{4\alpha_s(Q^2)}{3r} + br,
\end{equation}
where $b$ is a constant and $\alpha_s(Q^2)$ is the running coupling of the strong interaction. The other three terms are spin dependent.
\begin{equation}\label{eq:interactionSS}
 H^{\text{SS}} = \frac{32\pi\alpha_s(Q^2)}{9m\bar{m}}\tilde{\delta}_\sigma(\bm{r}) \bm{s}\cdot\bm{\bar{s}}
\end{equation}
is the spin-spin contact hyperfine potential, where $\bm{s}$ and $\bm{\bar{s}}$ are the spin of the quark and antiquark respectively, and $\tilde{\delta}_\sigma(\bm{r}) = (\frac{\sigma}{\sqrt{\pi}})^3 \text{e}^{-\sigma^2 r^2}$ with $\sigma$ being a parameter.
\begin{equation}\label{eq:interactionT}
 H^{\text{T}} = \frac{4\alpha_s(Q^2)}{3m\bar{m}} \frac{1}{r^3}\left( 3\frac{(\bm{s}\cdot\bm{r})(\bm{\bar{s}}\cdot\bm{r})}{r^2} - \bm{s}\cdot\bm{\bar{s}} \right)
\end{equation}
is the tensor potential. $H^{\text{SO}}$ is the spin-orbit interaction potential
\begin{equation}
\label{eq:interactionSO}
 H^{\text{SO}}  =  \frac{\bm{S}\cdot\bm{L}}{2}\left[ (\frac{1}{2m^2} + \frac{1}{2\bar{m}^2})(\frac{4\alpha_s(Q^2)}{3r^3} - \frac{b}{r})+ \frac{8\alpha_s(Q^2)}{3m\bar{m}r^3} \right]
\end{equation}
where $\bm{S} = \bm{s} + \bm{\bar{s}}$, and $\bm{L}$ is the orbital angular momentum of the quark and antiquark system.

In equations (\ref{eq:interactionSI})$\sim$(\ref{eq:interactionSO}), the running coupling takes the following form \cite{Godfrey1985},
\begin{equation}\label{eq:runningAlphas}
 \alpha_s(Q^2) = \frac{4\pi}{\beta \log(\text{e}^{\frac{4\pi}{\beta\alpha_0}} + \frac{Q^2}{\Lambda^2_{\text{QCD}}})},
\end{equation}
where $\Lambda_{\text{QCD}}$ is the energy scale below which nonperturbative effects take over, $\beta = 11-\frac{2}{3}N_f$ with $N_f$ being the flavor number, $Q$ is the momentum transfer, and $\alpha_0$ is a constant. Eq. (\ref{eq:runningAlphas}) approaches the one loop running form of QCD at large $Q^2$ and saturates at low $Q^2$.

The potentials containing $\frac{1}{r^3}$, Eq. (\ref{eq:interactionT}) and Eq. (\ref{eq:interactionSO}), are divergent. A cutoff $r_c$ is introduced to eliminate the divergence, i.e. $\frac{1}{r^3} \to \frac{1}{r^3_c}$ for $r \leq r_c$. Herein $r_c$ is a parameter to be fixed by observables.

$N_f$ and $\Lambda_{\text{QCD}}$ are chosen according to QCD estimation. The other parameters, $m_c$, $\alpha_0$, $b$, $\sigma$ and $r_c$, are tuned to fit the mass spectra of charmonium. Herein we use the same parameters  as Ref. \cite{Ding2025}, which yields resonable charmonium mass spectra and two-photon transition form factors of $\eta_{c}$ and $\chi_{c0}$

The electromagnetic form factor of pseudoscalar meson is defined as
\begin{equation}\label{eq:EFF1S0}
N_c\cdot\langle M(P_2)| j^\mu(0)|M(P_1)\rangle = F_p(Q^2)\cdot(P_1 + P_2)^\mu,
\end{equation}
where $|M(P_i)\rangle$ is the meson state with 4-momentum $P_i$, $j^\mu(0) = \bar{\psi}(0)\gamma^\mu \psi(0)$ is the vector current with $\psi(x)$ being the quark field operator, $F_p(Q^2)$ is the form factor, $Q^2$ is the square of the momentum transfer, $N_c$ is the color number.

Traditionally the meson is expressed by the mock meson state (MMS) in quark model \cite{Godfrey1985,Lakhina2006,Sun2023},
\begin{eqnarray}\nonumber
 |M(p) \rangle &=& \sqrt{\frac{2E_{\bm{p}}}{N_c}}\chi^{\bm{SM_S}}_{\bm{s\bar{s}}} \int\frac{d^3\bm{k} d^3\bm{\bar{k}}}{(2\pi)^3}\varphi_M\left(\bm{k}_r\right) \\\label{eq:oldMockMeson}
 && \cdot\delta^{(3)}(\bm{k}+\bm{\bar{k}}-\bm{p}) b^\dag_{\bm{ks}} d^\dag_{\bm{\bar{k}\bar{s}}}| 0 \rangle,
 \end{eqnarray}
where $\bm{p}$ and $E_{\bm{p}} =\sqrt{M^2 + \bm{p}^2}$ are the momentum and energy of the meson, $M$ is the meson mass, $N_c$ is the color number. $b^\dag_{\bm{ks}}$ and $d^\dag_{\bm{\bar{k}\bar{s}}}$ are the creation operator of the quark and antiquark respectively. $\chi^{\bm{SM_S}}_{\bm{s\bar{s}}}$ is the spin wave function, with $\bm{S}$ being the total spin and $\bm{M_S}$ its z-projection. $\varphi_M\left(\bm{k}_r\right)$ is the wave function in momentum space, $\bm{k}_r=\frac{\bar{m}\bm{k} - m\bm{\bar{k}}}{m+\bar{m}}$ is the relative momentum between the quark and antiquark. Herein we use the same symbol $\varphi$ for the wave functions in coordinate space and momentum space. In equation (\ref{eq:oldMockMeson}) a Clebsch-Gordan coefficient is indicated to combine the spin and orbital angular momentum into the total angular momentum of the meson.

A relativized mock meson state (RMMS) is proposed in Ref. \cite{Ding2025},
\begin{eqnarray}\nonumber
|M(p) \rangle &=& \sqrt{\frac{2E_{\bm{p}}}{N_c}} \sum_{\bm{s,\bar{s}}} \int\frac{d^3\bm{k} d^3\bm{\bar{k}}}{(2\pi)^6}\frac{1}{\sqrt{2E_{\bm{k}}}} \frac{1}{\sqrt{2\bar{E}_{\bar{\bm{k}}}}}\\\label{eq:RelativisedMockMeson}
&&\times \bar{u}(\bm{k},\bm{s}) \Gamma_M  v(\bar{\bm{k}},\bar{\bm{s}}) b^\dag_{\bm{ks}} d^\dag_{\bm{\bar{k}\bar{s}}} | 0 \rangle \cdot \epsilon.
\end{eqnarray}
where ${u}(\bm{k},\bm{s})$ and $v(\bar{\bm{k}},\bar{\bm{s}})$ are the Dirac spinors of the quark and antiquark respectively, $\Gamma_M$ is the meson Bethe-Salpeter amplitude (BSA), $\epsilon$ is the polarization tensor. For pseudoscalar meson the BSA is generally composed of 4 terms \cite{Chen2017}, however, we only employ the main term for simplicity, i.e.
\begin{equation}\label{eq:BSA1S0}
 \Gamma_P=\Gamma_P(k_r,p) = \gamma_5 \phi_P(k_r,p),
\end{equation}
where $k_r$ and $p$ are the relative and total 4-momentum, $\gamma_5$ is the Dirac matrix, $\phi_P(k_r,p)$ is a Lorentz scalar wave function. In principle, $\phi_P(k_r,p)$ should be solved from a covariant dynamical equation, e.g., the Bethe-Salpeter equation. However, in practice, we do not solve $\phi_P(k_r,p)$ rigorously. We relate $\phi_P(k_r,p)$ to the non-relativistic wave functions by requiring that Eq. (\ref{eq:RelativisedMockMeson}) approaches Eq. (\ref{eq:oldMockMeson}) in the non-relativistic limit, which leads to the following relation
\begin{equation}\label{eq:Relate1S0}
\phi_P(k_r,p)\!\! =\!\! \frac{\varphi_P\left(|\bm{k}_r|\right)}{\sqrt{8\pi}} (2\pi)^3\delta^{(3)}(\bm{k}+\bm{\bar{k}}-\bm{p}).                                                                                                                                                                                                                                                                                                                                                                                                                                                                                                                                                                                                                                                 \end{equation}
Equations (\ref{eq:RelativisedMockMeson})-(\ref{eq:Relate1S0}) describe the two-photon transition form factors of the heavy quarkonium reasonably well \cite{Ding2025}.

\begin{table}[!t]
\caption{\label{tab:BSA} The BSAs of the mesons used in our calculation. $ \bm{1} = \bm{1}_{4\times 4}$ is the unit matrix in Dirac space.}
\begin{tabular}{c|c|c|c|c|c|c}
\hline
\makecell[c]{\vspace{0.2em}} meson & $\eta_c$ & $\chi_{c0}$& $J/\psi$ & $\chi_{c1}$ & $h_c$ & $\chi_{c2}$ \\\hline
\makecell[c]{\vspace{0.4em}} $J^{PC}$ & $0^{-+}$ & $0^{++}$& $1^{--}$ & $1^{++}$ & $1^{+-}$ & $2^{++}$ \\
\hline
$\Gamma_M$ & $\gamma_5 \phi_P$ & $\bm{1}\!\cdot\! \phi_S$ & $\gamma^\mu \phi_V$ & $\gamma^\mu\gamma_5 \phi_{A1}$ & $k_{r}^{\mu}\gamma_5 \phi_{A2}$ & $(\gamma^\mu k_{r}^{\nu}\!\! +\!\!\gamma^\nu k_{r}^{\mu})\phi_T$\\
\hline
\end{tabular}
\end{table}

The RMMS of $J^{PC} = 0^{-+}$ and $0^{++}$ meson have been established in Ref. \cite{Ding2025}. In this paper we extend the definition of RMMS to $J^{PC} = 1^{--}$, $1^{++}$, $1^{+-}$ and $2^{++}$ meson, and apply them to calculate the single-quark electromagnetic form factors of the charmonium. The meson BSAs are listed in Table \ref{tab:BSA}. We put the derivations of the non-relativistic limit of the RMMS in \ref{sec:appendixA}. Requiring Eq. (\ref{eq:RelativisedMockMeson}) approaches Eq. (\ref{eq:oldMockMeson}) in the non-relativistic limit, we get the following form of the Lorentz scalar wavefunctions,
\begin{eqnarray}\label{eq:Relate3P0}
 \phi_S(k_r,p)\!\! &=&\!\!\! \frac{2E_{\bm{k}} \bar{E}_{\bar{\bm{k}}}}{E_{\bm{k}} + \bar{E}_{\bar{\bm{k}}}}\frac{\varphi_S\left(|\bm{k}_r|\right)}{\sqrt{8\pi}|\bm{k}_r|} (2\pi)^3\delta^{(3)}(\bm{k}\!+\!\bm{\bar{k}}\!-\!\bm{p}),\\\nonumber
\phi_V(k_r,p)\!\! &=&\!\! \sqrt{\frac{2E_{\bm{k}} \bar{E}_{\bar{\bm{k}}}}{E_{\bm{k}}\bar{E}_{\bar{\bm{k}}} +m\bar{m} +\bm{k}^2_r/3}} \\\label{eq:Relate3S1}
&&\times \frac{\varphi_V\left(|\bm{k}_r|\right)}{\sqrt{8\pi}} (2\pi)^3\delta^{(3)}(\bm{k}+\bm{\bar{k}}-\bm{p}),\\\nonumber
\phi_{A1}(k_r,p)\!\! &=&\!\! \sqrt{\frac{4E_{\bm{k}} \bar{E}_{\bar{\bm{k}}} \bm{k}^2_r }{ 3m\bar{m} (E_{\bm{k}}\bar{E}_{\bar{\bm{k}}} -m\bar{m} +\bm{k}^2_r/3)}} \\\label{eq:Relate3P1}
&&\hspace{-1.5em}\times \frac{\sqrt{3m\bar{m}}\varphi_{A1}\left(|\bm{k}_r|\right)}{\sqrt{8\pi}|\bm{k}_r|} (2\pi)^3\delta^{(3)}(\bm{k}+\bm{\bar{k}}-\bm{p}),\\\label{eq:Relate1P1}                                                                                                                                                                                                                                                                                                                                                                                                                                                                                                                                                                                                                                                   \phi_{A2}(k_r,p)\!\! &=&
\frac{\sqrt{3}\varphi_{A2}\left(|\bm{k}_r|\right)}{\sqrt{8\pi}|\bm{k}_r|} (2\pi)^3\delta^{(3)}(\bm{k}+\bm{\bar{k}}-\bm{p}),\\\nonumber
\phi_{T}(k_r,p)\!\! &=&\!\! \sqrt{\frac{2E_{\bm{k}} \bar{E}_{\bar{\bm{k}}}}{E_{\bm{k}}\bar{E}_{\bar{\bm{k}}} +m\bar{m} +\bm{k}^2_r/3}} \\\label{eq:Relate3P2}
&&\times \frac{\sqrt{3}\varphi_{T}\left(|\bm{k}_r|\right)}{\sqrt{32\pi}|\bm{k}_r|} (2\pi)^3\delta^{(3)}(\bm{k}+\bm{\bar{k}}-\bm{p}).                                                                                                                                                                                                                                                                                                                                                                                                                                                                                                                                                                                                                     \end{eqnarray}

Put Eq. (\ref{eq:oldMockMeson}) or Eq. (\ref{eq:RelativisedMockMeson}) into Eq. (\ref{eq:EFF1S0}), we get the single-quark electromagnetic form factor of pseudoscalar meson. This computational scheme is equally applicable to higher spin mesons. For a particle with integral spin $J$, there are $2J+1$ independent form factors \cite{Lorce2009},
\begin{eqnarray}\nonumber
&&N_c\cdot\langle M(P_2)| j^\mu(0)|M(P_1)\rangle = \\\label{eq:EFFJ}
&&(-1)^J \epsilon^*_{\alpha'_1\cdots \alpha'_J}\left[(P_1+P_2)^\mu\sum_{(k,J)} F_{2k+1}(Q^2) \right. \\\nonumber
&&+\left. (g^{\mu\alpha_J}q^{\alpha'_J} - g^{\mu\alpha'_J}q^{\alpha_J})\sum_{(k,J-1)} F_{2k+2}(Q^2)  \right]\epsilon_{\alpha_1\cdots \alpha_J},
\end{eqnarray}
where $F_i$ are the form factors, $g^{\mu\nu}$ is the Minkowski metric, $q = P_2 - P_1$ is the 4-momentum transfer, $Q^2 = -q^2$, $\epsilon_{\alpha_1\cdots \alpha_J}$ is the polarization tensor. The sum stands for
\begin{equation}
 \sum_{(k,J)} \equiv \sum_{k=0}^J \left[\prod_{i=1}^k(-\frac{q^{\alpha_i} q^{\alpha'_i}}{2M^2}) \prod_{i=k+1}^J  g^{\alpha_i\alpha'_i} \right].
\end{equation}
The multipole form factors are related to the covariant vertex functions $F_i$ by
\begin{eqnarray}\nonumber
&& \sum_{m=t}^J (-1)^{m+t} \frac{\tau^{m-t} (\text{C}_m^t)^2 }{\tilde{\text{C}}_{4m-1}^{2m+2t-1}} G_{E2m} =\\
&& \sum_{k=0}^t (1+\tau)^k \text{C}_{J-k}^{J-t} [F_{2k+1} - \frac{1-\delta_{k0}}{1+\tau} F_{2k}],\\\nonumber
&& \sum_{m=t}^{J-1} (-1)^{m+t} (m+1)\frac{\tau^{m-t} (\text{C}_m^t)^2 }{\tilde{\text{C}}_{4m+1}^{2m+2t+1}} G_{M2m+1} =\\
&&(t+1) \sum_{k=0}^t (1+\tau)^k \text{C}_{J-k-1}^{J-t-1} F_{2k+2},
\end{eqnarray}
where $G_{Em}$ and $G_{Mm}$ are the multipole electric and magnetic form factors, $\tau = \frac{Q^2}{4M^2}$, $\text{C}_m^t = \frac{m!}{t!(m-1)!}$, $\tilde{\text{C}}_m^t = \frac{m!!}{t!!(m-1)!!}$. Note that the multipole form factors differ from Ref. \cite{Lorce2009} by a factor $\sqrt{1+\tau}$, and is consistent with Ref. \cite{Arnold1980} in the $J=1$ case. The electric moment of order $l$, $Q_l$, in natural unit of $\frac{e}{M^l}$ is given by \cite{Lorce2009}
\begin{equation}
Q_l = \frac{(l!)^2}{2^{l}} G_{El}(0).
\end{equation}
The magnetic moment of order $l$, $\mu_l$, in natural unit of $\frac{e}{2M^l}$ is given by \cite{Lorce2009}
\begin{equation}
\mu_l = \frac{(l!)^2}{2^{l-1}} G_{Ml}(0).
\end{equation}

In this paper we consider 12 mesons as follows:
$\eta_c(1S)$, $\eta_c(2S)$, $\chi_{c0}(1P)$, $\chi_{c0}(2P)$, $J/\psi(1S)$, $J/\psi(2S)$, $\chi_{c1}(1P)$, $\chi_{c1}(2P)$, $h_c(1P)$, $h_c(2P)$, $\chi_{c2}(1P)$ and $\chi_{c2}(2P)$, and the details of calculating the single-quark form factors are put in \ref{sec:appendixB}. The electric and magnetic multipole moments of order $m$ are designated as $G_{Em}$ and  $G_{Mm}$ respectively. For $J=0$ mesons ($\eta_c(1S)$, $\eta_c(2S)$, $\chi_{c0}(1P)$, $\chi_{c0}(2P)$), the covariant form factor equals $G_{E0}$, e.g., $F_{\eta_c(1S)} = G_{E0,\eta_c(1S)}$. We also adopt the following symbols for the low-order multipoles: electric charge form factor $G_C = G_{E0}$, magnetic dipole form factor $G_M = G_{M1}$, electric quadrupole form factor $G_Q = G_{E2}$. For higher-order multipole moments, we employ a general notation, e.g., $G_{M3}$ for magnetic octupole form factor and $G_{E4}$ for electric hexadecapole form factor.

\begin{figure}[!t]
\centering
 \includegraphics[width=0.48\textwidth]{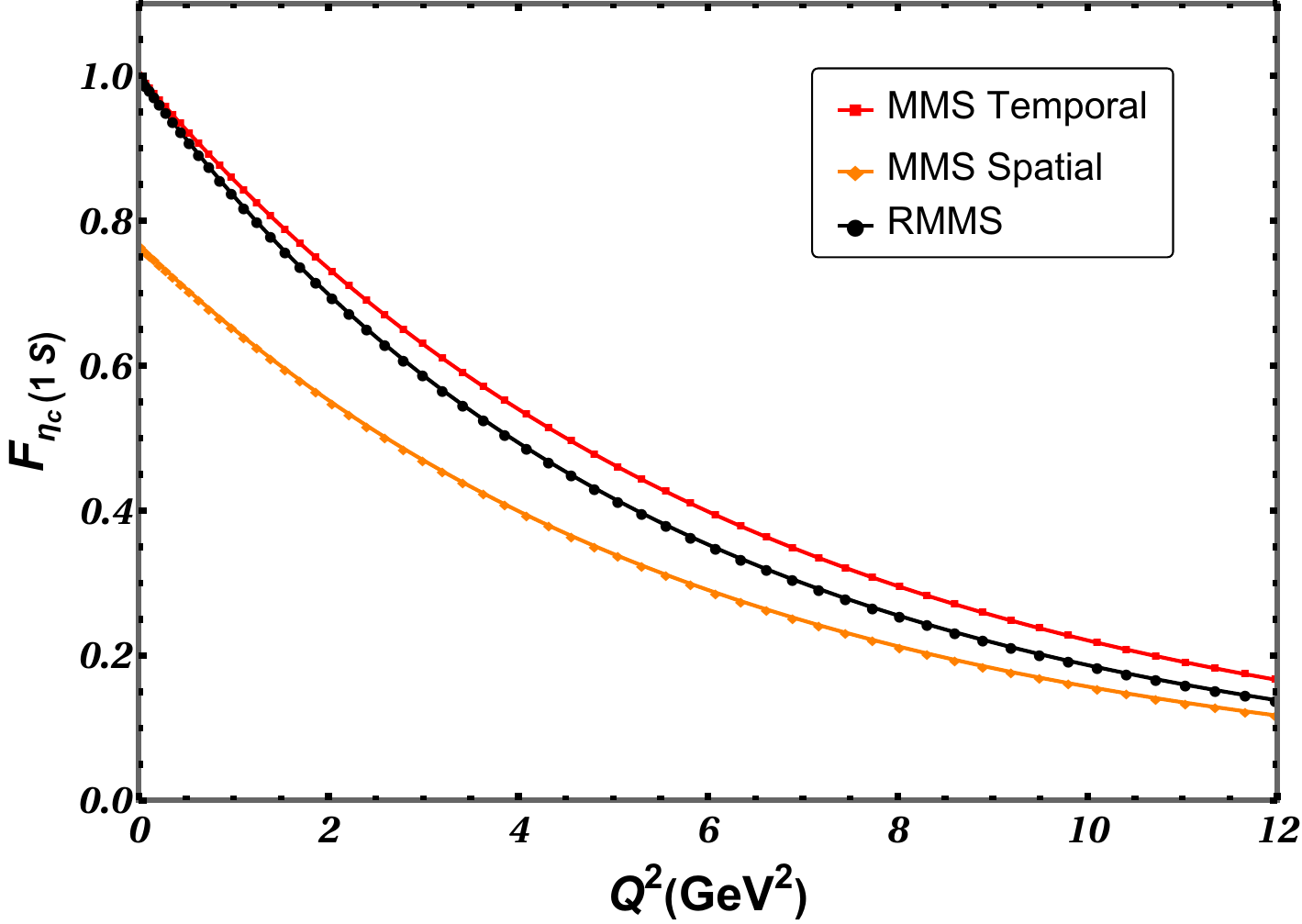}
  \includegraphics[width=0.48\textwidth]{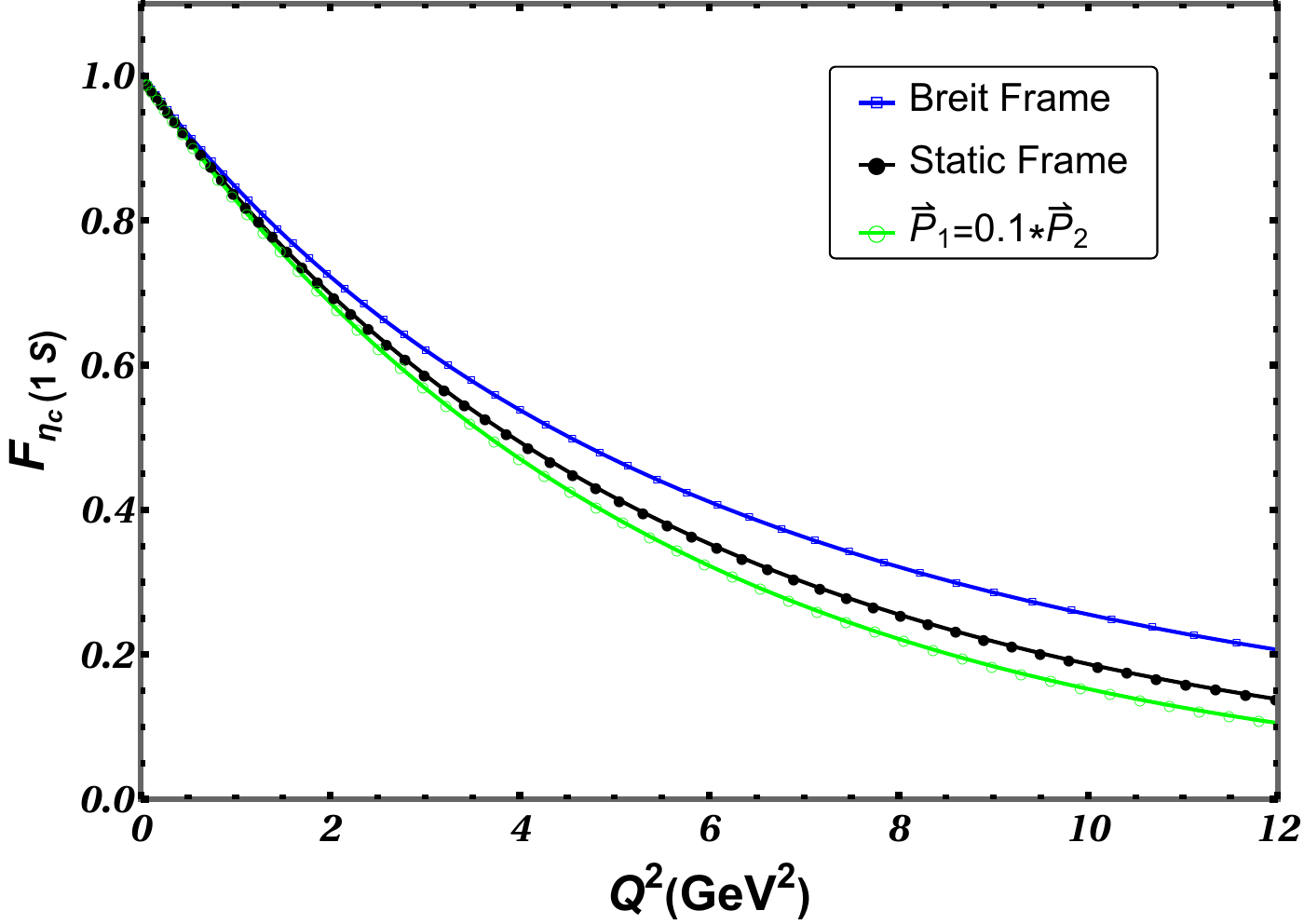}
 \caption{\label{fig:effetac1SLiF} Upper panel: Comparison of $F_{\eta_c(1S)}$ computed in the static frame using: (top) MMS with the temporal component current $j^0$, (middle) RMMS, (bottom) MMS with the spacial component current $j^i$ (i=1,2,3). Lower panel shows the results of $F_{\eta_c(1S)}$ in different reference frame using RMMS. From top to bottom, the Breit frame, the static frame and a frame where $\bm{P}_1 = 0.1*\bm{P}_2$ holds.}
\end{figure}

Before discussing the results, let's first emphasize the computational errors arising from the violation of Lorentz covariance. Using MMS, Eq. (\ref{eq:oldMockMeson}), the errors of the form factor due to losing Lorentz covariance manifest in two aspects: dependence on the Lorentz index of $j^\mu(0)$ and dependence on the reference frame \cite{Lakhina2006}. Specifically, there is a discrepancy between the spatial component ($\mu=1,2,3$) and temporal component ($\mu=0$) results, and the outcomes vary when computed in different reference frames. Although the RMMS, Eq. (\ref{eq:RelativisedMockMeson}), appears Lorentz covariant, the form factors still depend on the reference frame as the wavefunctions, Eq. (\ref{eq:Relate1S0})- Eq. (\ref{eq:Relate3P2}) aren't really Lorentz scalar functions. To investigate the frame dependence of the result, we express the momenta of the mesons in a general form,
\begin{equation}
 P_1 = (E_{\bm{P}_1}, \bm{P}_1),\quad P_2 = (E_{\bm{P}_2}, \bm{P}_2)
\end{equation}
where $\bm{P}_1$ and $\bm{P}_2$ are the 3-momenta of the mesons, $E_{\bm{P}_i} = \sqrt{M^2 + \bm{P}_i^2}\;(i=1,2)$ are the energies. The 4-momentum transfer is $q = P_2 - P_1$, and $Q^2 = -q^2$. Two specific reference frames will be employed in our calculation,
\begin{eqnarray}
\text{Breit frame}: &\bm{P}_1 = -\bm{P}_2,\\
\text{Static frame}: &\bm{P}_1 = 0.
\end{eqnarray}
In the Breit frame, the finnal meson and the initial meson move in opposite direction. In the static frame, the initial meson is at rest.

A comparison of the results of $F_{\eta_c(1S)}$ using MMS and RMMS is presented in the upper panel of Fig. \ref{fig:effetac1SLiF}. The result from MMS with spacial compoment current is lower than that with temporal compoment current, which is in agreement with Ref. \cite{Lakhina2006}. The result from RMMS is an intermediate between the other two. At lower $Q^2$ region it nears the ``MMS temporal" result and at larger $Q^2$ region it trends toward the ``MMS spacial" result. The lower panel of Fig. \ref{fig:effetac1SLiF} demonstrates how the result depends on the choice of the reference frame, with the example of $F_{\eta_c(1S)}$ using RMMS. In a reference frame where the finnal meson and the initial meson move in opposite direction, the form factor is larger than that in the static frame. The form factor in the Breit frame provides an upper bound of the results. In a reference frame where the finnal meson and the initial meson move in the same direction, the form factor is lower than that in the static frame, which is illustrated in the case of a frame where $\bm{P}_1 = 0.1*\bm{P}_2$ holds. In the following we adopt the RMMS result as our prediction, taking the result in the static frame  as the central value and treating the difference of the results between the Breit frame and the static frame as the error.

\section{The results}\label{sec:resultsCC}

Our results of the single-quark electric form factors of $\eta_c(1S)$, $\eta_c(2S)$, $\chi_{c0}(1P)$ and $\chi_{c0}(2P)$, the electric, magnetic dipole and electric quadrupole form factors of $J/\psi(1S)$, $J/\psi(2S)$, $\chi_{c1}(1P)$, $\chi_{c1}(2P)$, $h_c(1P)$, $h_c(2P)$, $\chi_{c2}(1P)$ and $\chi_{c2}(2P)$, the magnetic octupole form factor and electric hexadecapole form factors of $\chi_{c2}(1P)$ and $\chi_{c2}(2P)$ are presented sequentially. We compare our results with those of the lattice quantum chromodynamics (LQCD), the Dyson-Schwinger equation (DSE) and the basis light front quantization (BLFQ) approach whence available. These results can be directly read from Fig. \ref{fig:effetac1S2S} to Fig. \ref{fig:effchic21P2Pb}. Most of our results agrees with the other results, and there are slight discrepancies in a few cases. In the following, we explain and discuss the results step-by-step.

\subsection{$\eta_c(1S)$ and $\eta_c(2S)$}\label{subsec:etac}

\begin{figure}[!t]
\centering
 \includegraphics[width=0.48\textwidth]{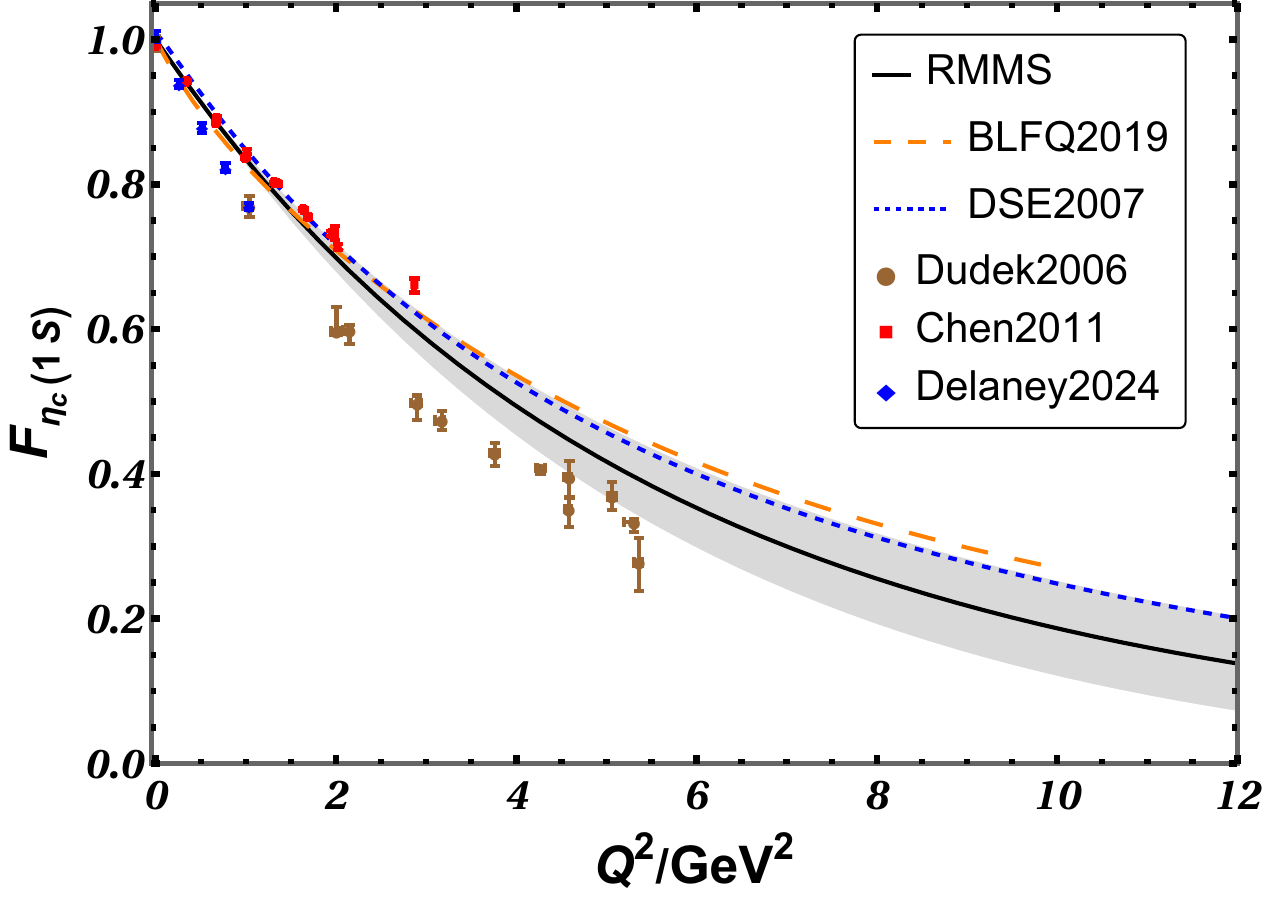}
  \includegraphics[width=0.48\textwidth]{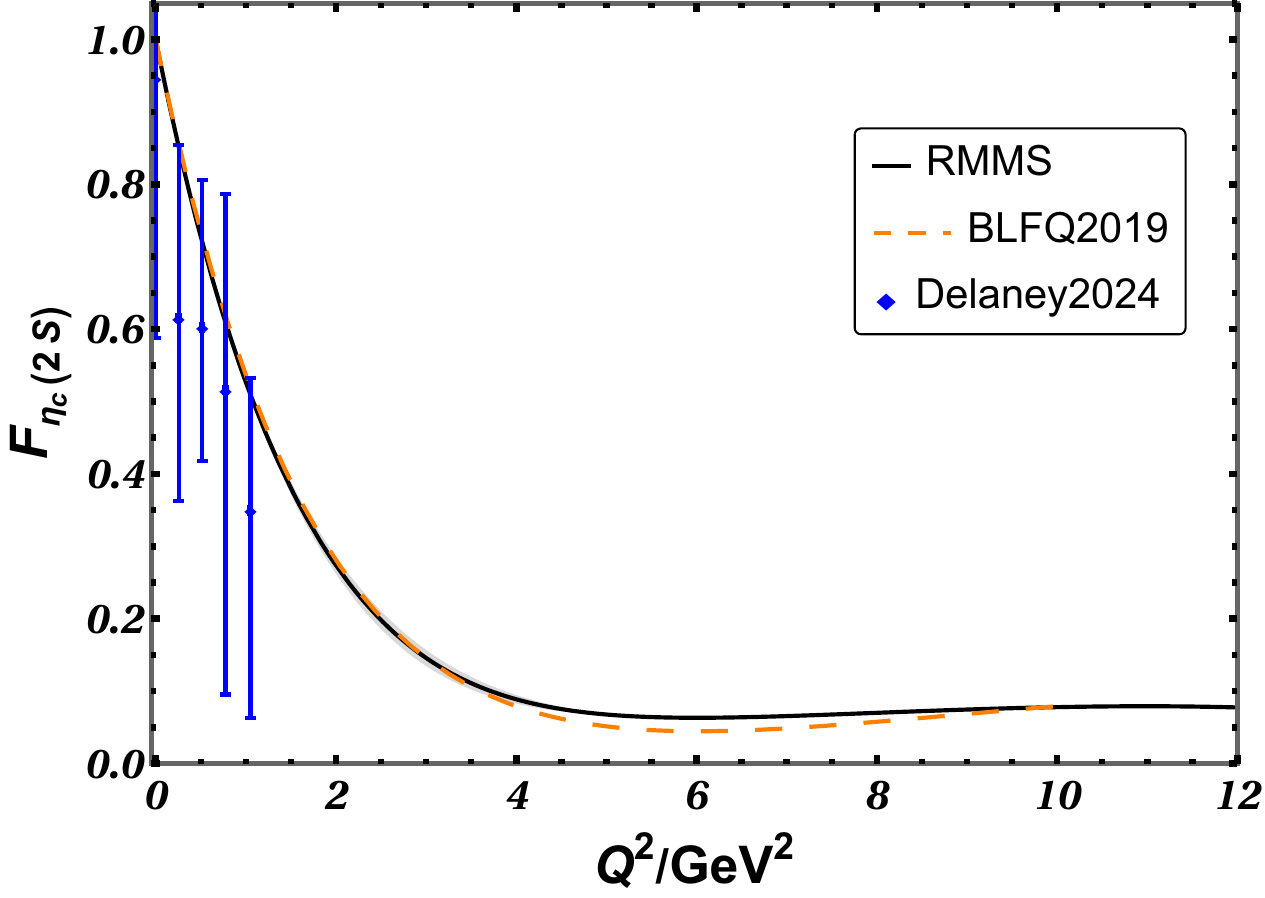}
 \caption{\label{fig:effetac1S2S} Single-quark electromagnetic form factor of $\eta_c(1S)$ (upper) and $\eta_c(2S)$ (lower). The solid black line (RMMS) is our result via the relativized mock meson state, and the shadow is the computational error. The dotted blue line (DSE2007) is the Dyson-Schwinger equation result \cite{Maris2007}. The dashed orange line (BLFQ2019) is the basis light front quantization result \cite{Adhikari2019}. The dots with error bars are the lattice QCD results: Dudek2006 \cite{Dudek2006}, Chen2011 \cite{Chen2011} and Delaney2024 \cite{Delaney2024}.}
\end{figure}

The single-quark electromagnetic form factor of $\eta_c(1S)$ is displayed in the upper panel of Fig. \ref{fig:effetac1S2S}. The solid black line (RMMS) is our prediction, and the shadow is the computational error. The upper limit of our result  (from the Breit frame) agrees well with the Dyson-Schwinger equation result (DSE2007 \cite{Maris2007}), and is consistent with the basis light front quantization result (BLFQ2019 \cite{Adhikari2019}) and the LQCD result (Chen2011 \cite{Chen2011}). Our result is slightly larger than the other two LQCD results, Dudek2006 \cite{Dudek2006} and Delaney2024 \cite{Delaney2024}. However, at larger $Q^2$ region, they remain consistent.

The single-quark electromagnetic form factor of $\eta_c(2S)$ is displayed in the lower panel of Fig. \ref{fig:effetac1S2S}. The computational error is significantly smaller than that of $\eta_c(1S)$, making the shaded region nearly invisible in the plot. Our result almost overlaps with the BLFQ2019 result, and is consistent with the Delaney2024 result, considering the large errors.

\subsection{$\chi_{c0}(1P)$ and $\chi_{c0}(2P)$}\label{subsec:chic0}

\begin{figure}[!t]
\centering
 \includegraphics[width=0.48\textwidth]{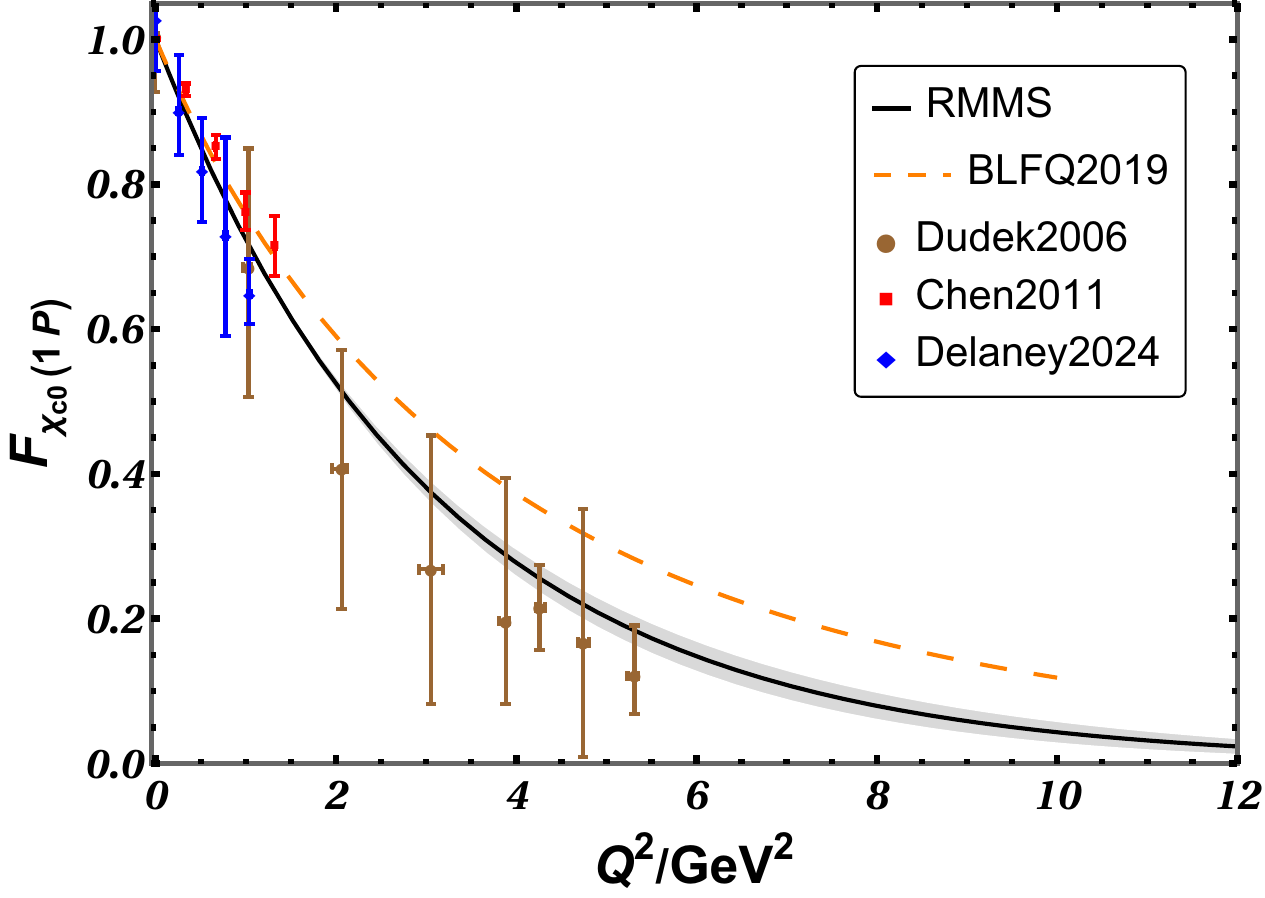}
  \includegraphics[width=0.48\textwidth]{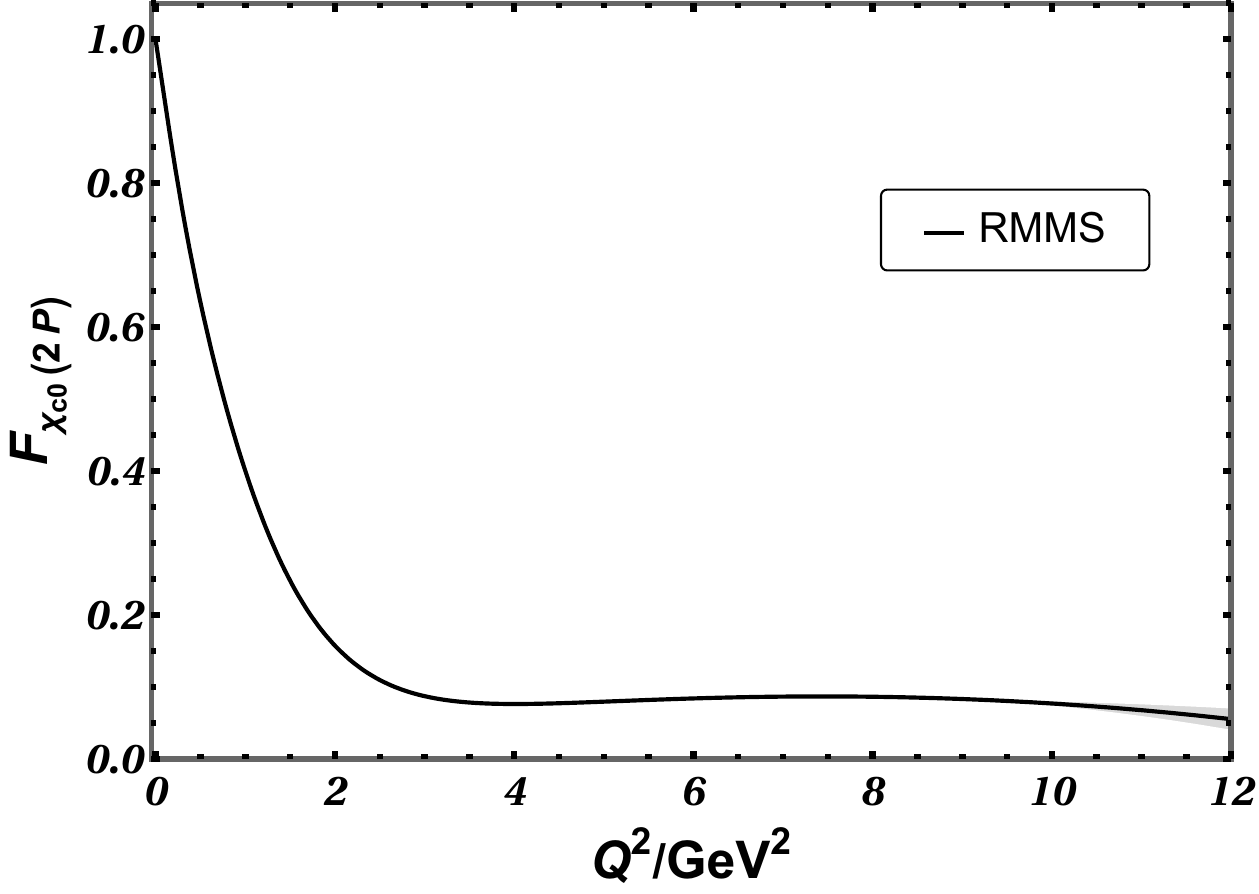}
 \caption{\label{fig:effchic01P2P} Single-quark electromagnetic form factor of $\chi_{c0}(1P)$ (upper) and $\chi_{c0}(2P)$ (lower). The caption is identical to that of FIG. \ref{fig:effetac1S2S}.}
\end{figure}

The single-quark electromagnetic form factor of $\chi_{c0}(1P)$ is displayed in the upper panel of Fig. \ref{fig:effchic01P2P}. Our result is consistent with all the three LQCD results: Dudek2006, Chen2011 and Delaney2024, and slightly lower than the BLFQ2019 result. Our prediction of  single-quark electromagnetic form factor of $\chi_{c0}(2P)$ is displayed in the lower panel of Fig. \ref{fig:effchic01P2P}, of which no comparable data is found.

\subsection{$J/\psi(1S)$ and $J/\psi(2S)$}\label{subsec:jpsi}

\begin{figure}[!t]
\centering
 \includegraphics[width=0.48\textwidth]{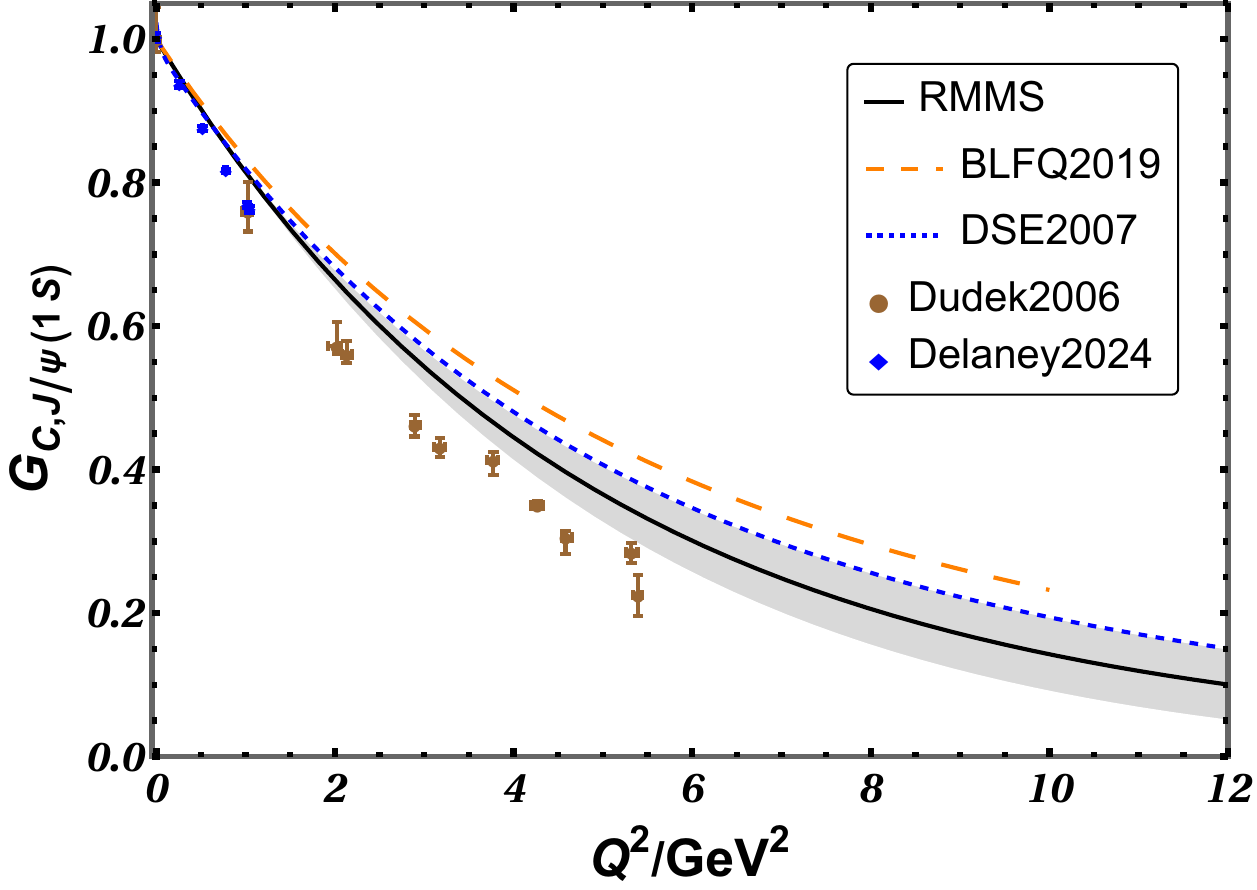}
 \includegraphics[width=0.48\textwidth]{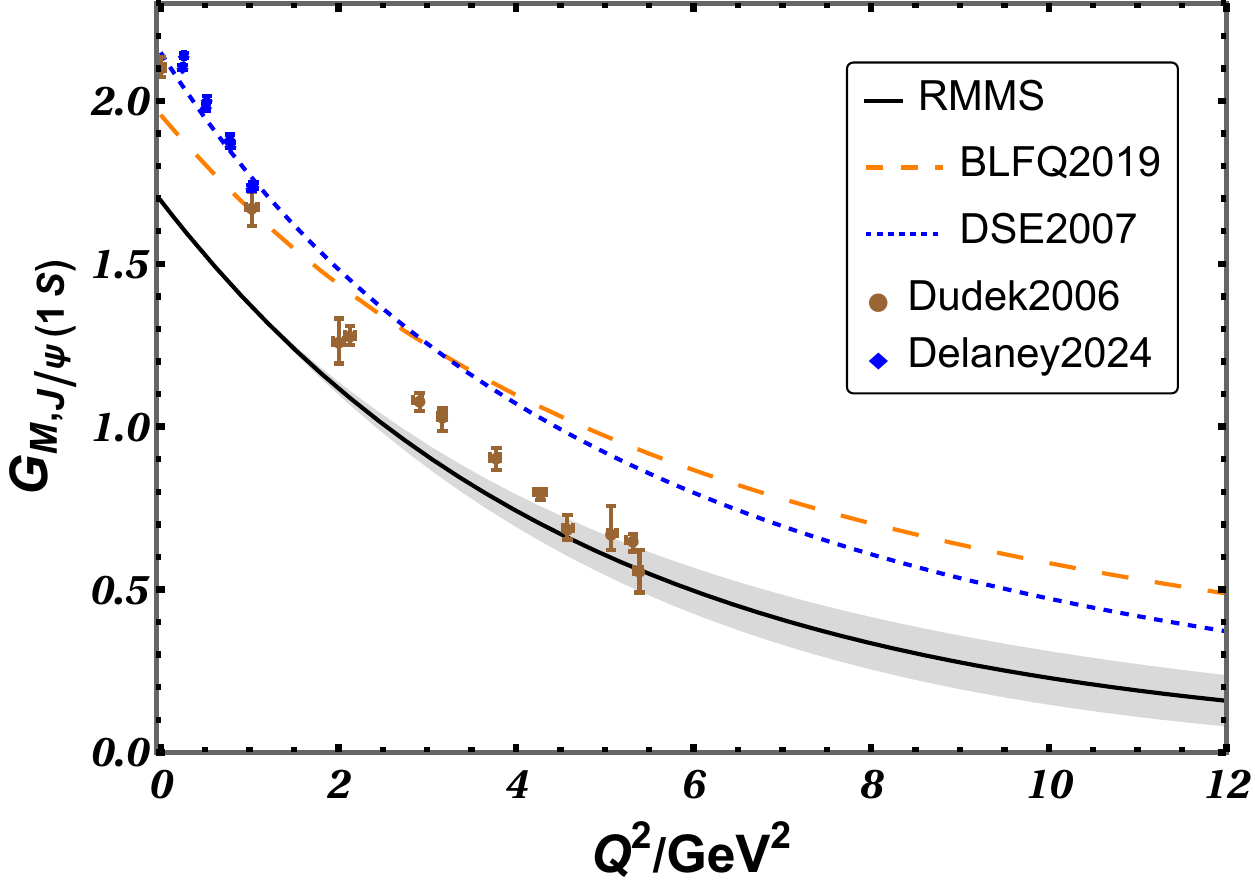}
 \includegraphics[width=0.48\textwidth]{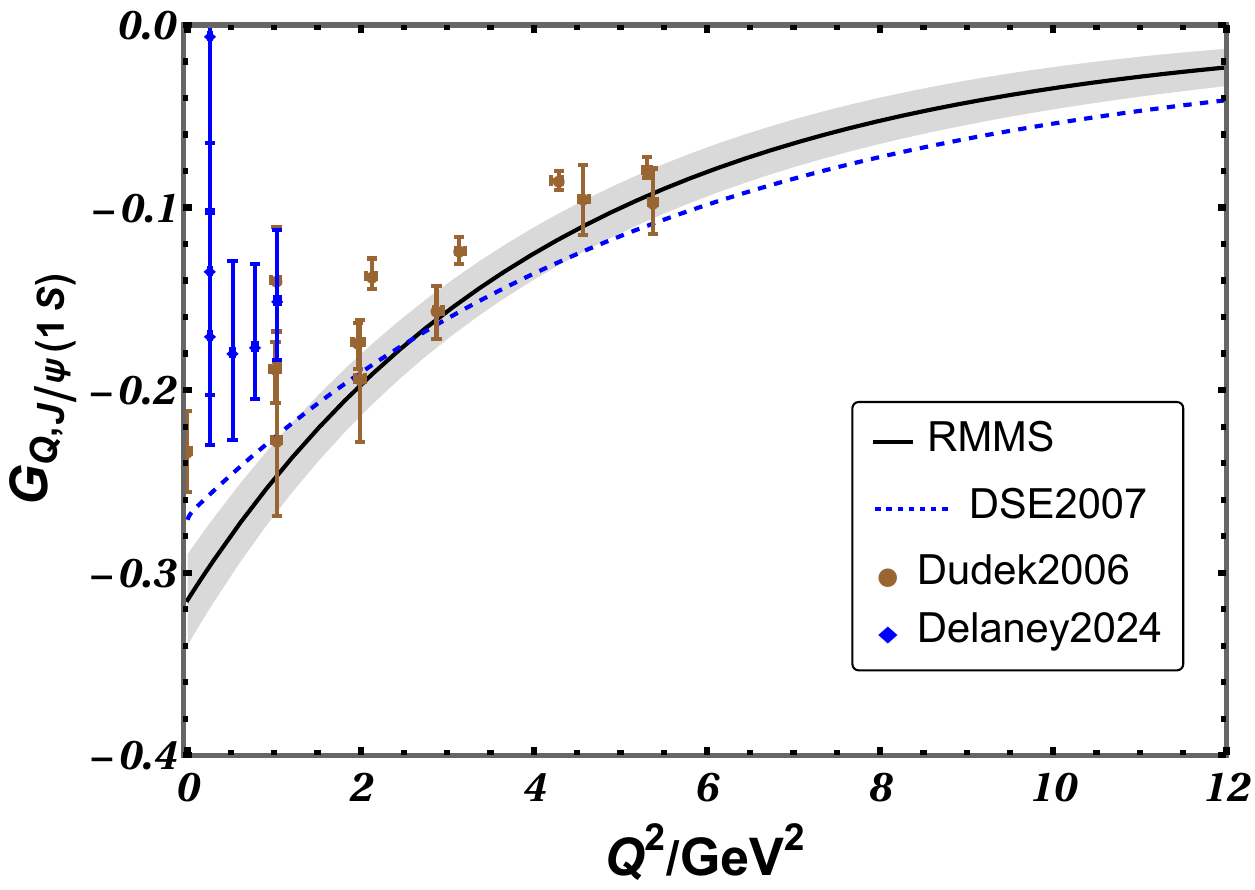}
 \caption{\label{fig:effJpsi1S} Single-quark electric charge (upper), magnetic dipole (middle) and electric quadrupole (lower) form factor of $J/\psi(1S)$. The caption is identical to that of FIG. \ref{fig:effetac1S2S}.}
\end{figure}

\begin{figure}[!t]
\centering
 \includegraphics[width=0.48\textwidth]{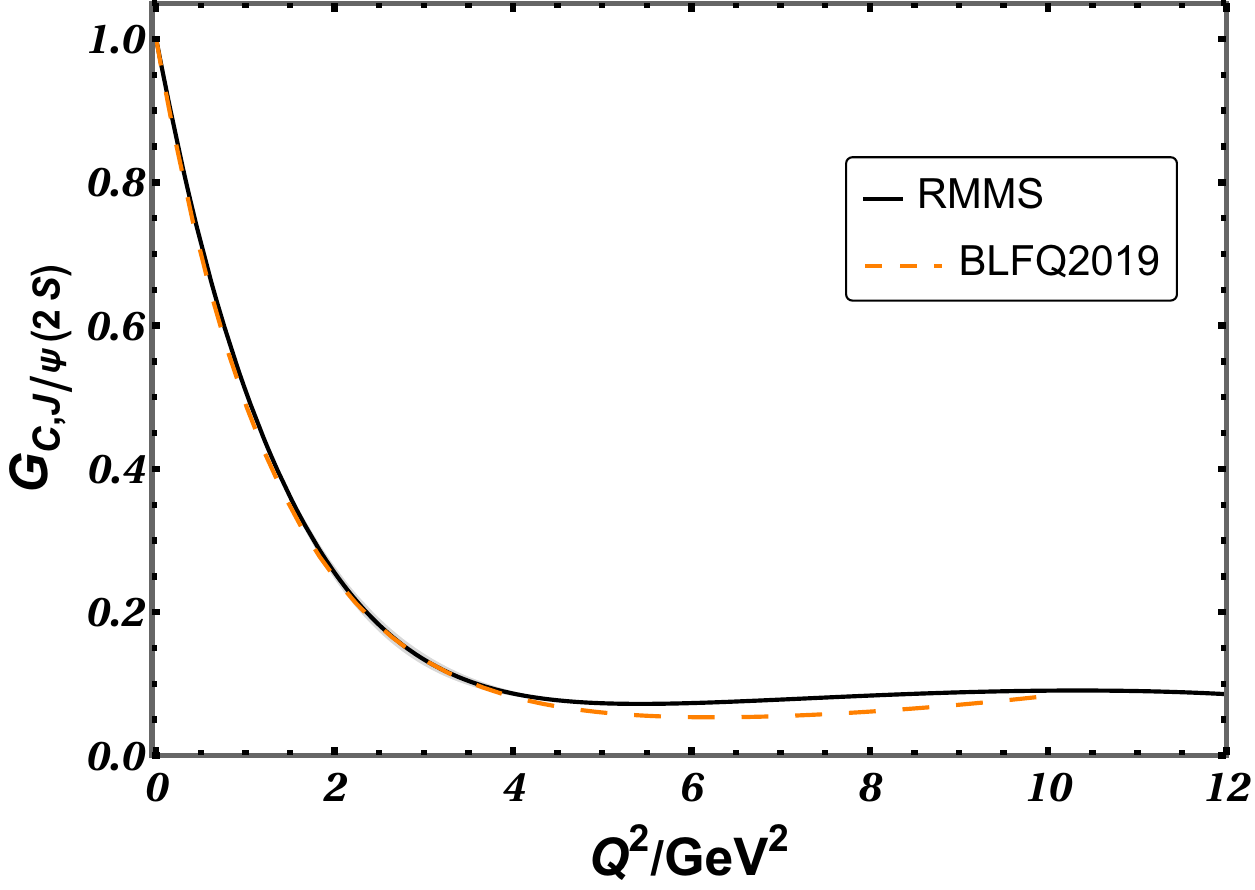}
 \includegraphics[width=0.48\textwidth]{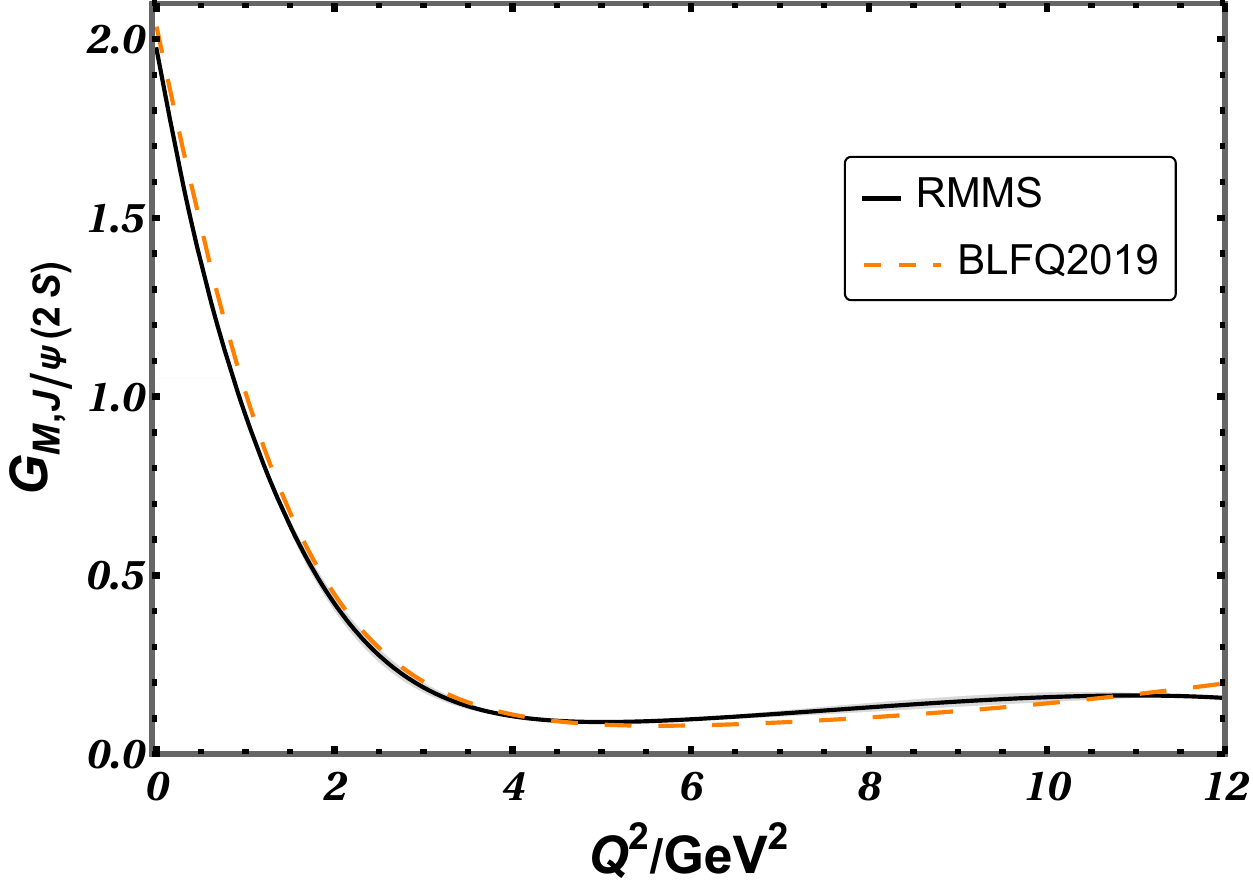}
 \includegraphics[width=0.48\textwidth]{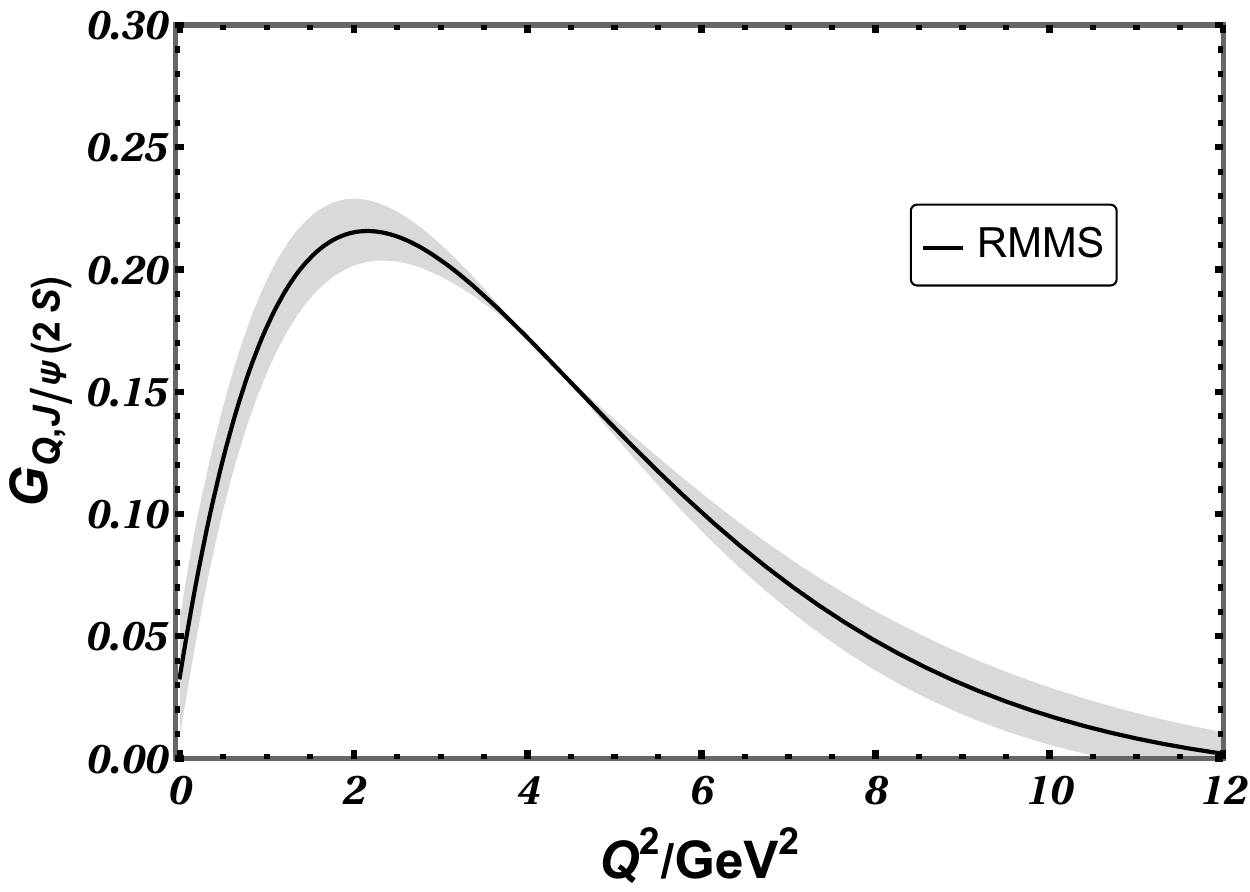}
 \caption{\label{fig:effJpsi2S} Single-quark electric charge (upper), magnetic dipole (middle) and electric quadrupole (lower) form factor of $J/\psi(2S)$. The caption is identical to that of FIG. \ref{fig:effetac1S2S}.}
\end{figure}

The single-quark electric form factor of $J/\psi(1S)$ is displayed in the upper panel of Fig. \ref{fig:effJpsi1S}. Again the upper limit of our result overlaps with the Dyson-Schwinger equation result (DSE2007 \cite{Maris2007}). Our result is slightly larger than the LQCD results: Dudek2006 \cite{Dudek2006} and Delaney2024 \cite{Delaney2024}, slightly lower than the BLFQ2019 \cite{Adhikari2019} result.

The single-quark magnetic dipole form factor of $J/\psi(1S)$ is displayed in the middle panel of Fig. \ref{fig:effJpsi1S}. Our result is lower than all the other results in the plot. This is the case where our result exhibits the greatest discrepancy with other results. Ref. \cite{Lakhina2006} also calculated $G_{M,J/\psi(1S)}$ via quark model using different parameters, and their result is lower than the LQCD result Dudek2006 \cite{Dudek2006}. This suggests that the discrepancy may originate from aspects beyond parameter tuning, and it may arise from the assumption of point-like quarks in the quark model. In the non-relativistic limit, the magnetic moment of a single quark in the meson is given by its Landé g-factor, $\mu_J = g\frac{e_q}{2m}J$, $e_q$ is the electric charge of the quark and
\begin{equation}\label{eq:gfactor}
 g = 1 + \frac{J(J+1)-L(L+1)+S(S+1)}{2J(J+1)}.
\end{equation}
Eq. (\ref{eq:gfactor}) shows that in the non-relativistic limit, the single-quark magnetic moment in unit of $\frac{e_q}{2m}$ in a $^3S_1$ meson is 2. The deviation of $G_{M,J/\psi}(0)$ from 2 is due to relativistic corrections. According to our calculation, $\mu_{1,J/\psi(1S)} = G_{M,J/\psi(1S)}(0)=1.70$, $\mu_{1,J/\psi(2S)} = G_{M,J/\psi(2S)}(0)=1.97$.

The single-quark electric quadrupole form factor of $J/\psi(1S)$ is displayed in the lower panel of Fig. \ref{fig:effJpsi1S}. Our result is in good agreement with the LQCD result Dudek2006 and the Dyson-Schwinger equation result DSE2007. A non-zero electric quadrupole moment indicates a non-spherical charge distribution, implying $J/\psi(1S)$ is not a pure S-wave state. Actually the vector RMMS, Eq. (\ref{eq:RMMS3S1}), contains a predominant S-wave component and a minor D-wave contribution. Keeping contributions through next-to-leading order, Eq. (\ref{eq:RMMS3S1}) is approximated as
\begin{eqnarray}\nonumber
 |V(p) \rangle &\approx &\sqrt{\frac{2E_{\bm{p}}}{N_c}} \sum\limits_{s\bar{s}}\int \frac{d^3\bm{k}}{(2\pi)^3}  \tilde{\varphi}_{V,{\text{mix}}}\left(\bm{k}_r\right) b^\dag_{\bm{ks}} d^\dag_{\bm{\bar{k}\bar{s}}}| 0 \rangle,\\\nonumber
 \tilde{\varphi}_{V,{\text{mix}}}\left(\bm{k}_r\right) &=& (1+\frac{\bm{k}^2}{12m\bar{m}})\times\tilde{\varphi}_{V,S}\left(\bm{k}_r\right)\\
 && + \frac{\bm{k}^2}{3\sqrt{2}m\bar{m}}\times\tilde{\varphi}_{V,D}\left(\bm{k}_r\right),
\end{eqnarray}
where $\tilde{\varphi}_{V,S}\left(\bm{k}_r \right)= \varphi_{V}\left(|\bm{k}_r| \right)\chi^{\bm{1M_J}}_{\bm{s\bar{s}}} Y^{00}$ represents the S-wave contribution, $\tilde{\varphi}_{V,D}\left(\bm{k}_r \right)= \varphi_{V}\left(|\bm{k}_r| \right)\chi^{\bm{1M_S}}_{\bm{s\bar{s}}} Y^{2,M_J-M_L}$ with a Clebsch-Gordan coefficient indicated represents the D-wave contribution. The dominant contribution to the electric quadrupole moment comes from the S-D wave interference term. According to our calculation, the single-quark electric quadrupole moments in unit of $\frac{e_q}{M^2}$ are $Q_{2,J/\psi(1S)} =  G_{Q,J/\psi(1S)}(0) = -0.315$ and $Q_{2,J/\psi(2S)} =G_{Q,J/\psi(2S)}(0) = 0.033$.

The single-quark electric form factor of $J/\psi(2S)$ is displayed in the upper panel of Fig. \ref{fig:effJpsi2S}, and the single-quark magnetic dipole form factor of $J/\psi(2S)$ is displayed in the middle panel of Fig. \ref{fig:effJpsi2S}. For the $Q^2$ range shown in the figure, both of our results are in very good agreement with the BLFQ2019 results. The single-quark electric quadrupole form factor of $J/\psi(2S)$ is displayed in the lower panel of Fig. \ref{fig:effJpsi2S}, of which no comparable data is found.

\subsection{$\chi_{c1}(1P)$ and $\chi_{c1}(2P)$}\label{subsec:chic1}

\begin{figure}[!t]
\centering
 \includegraphics[width=0.48\textwidth]{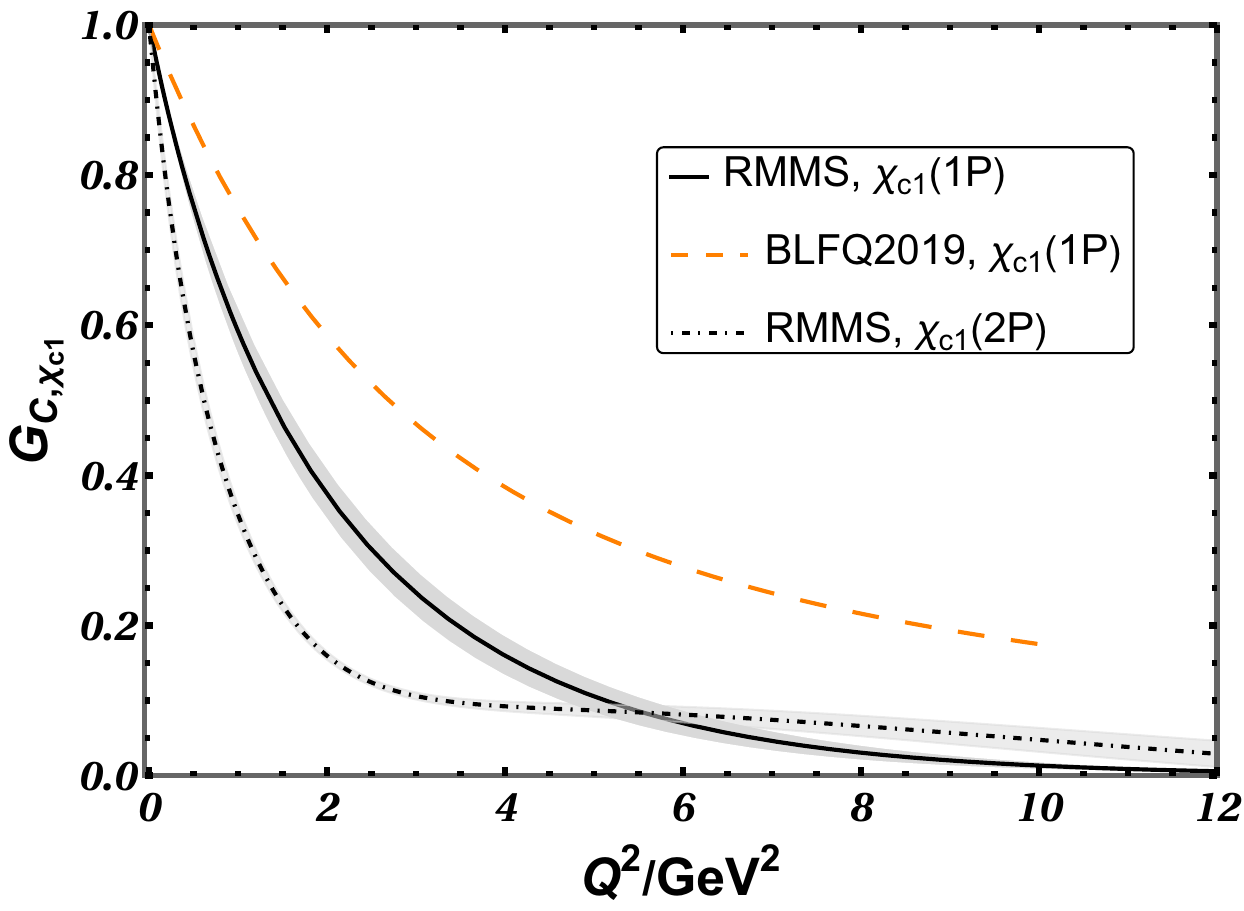}
 \includegraphics[width=0.48\textwidth]{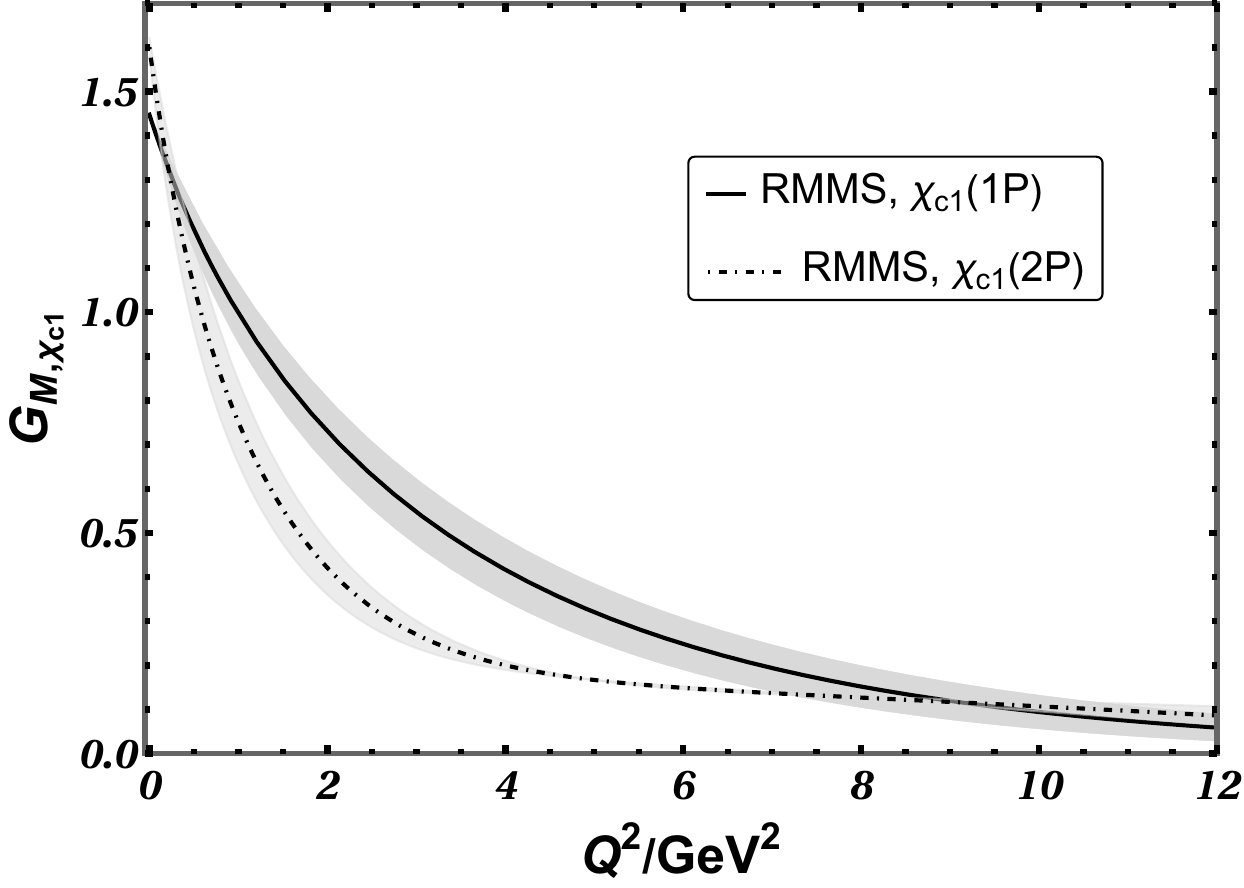}
 \includegraphics[width=0.48\textwidth]{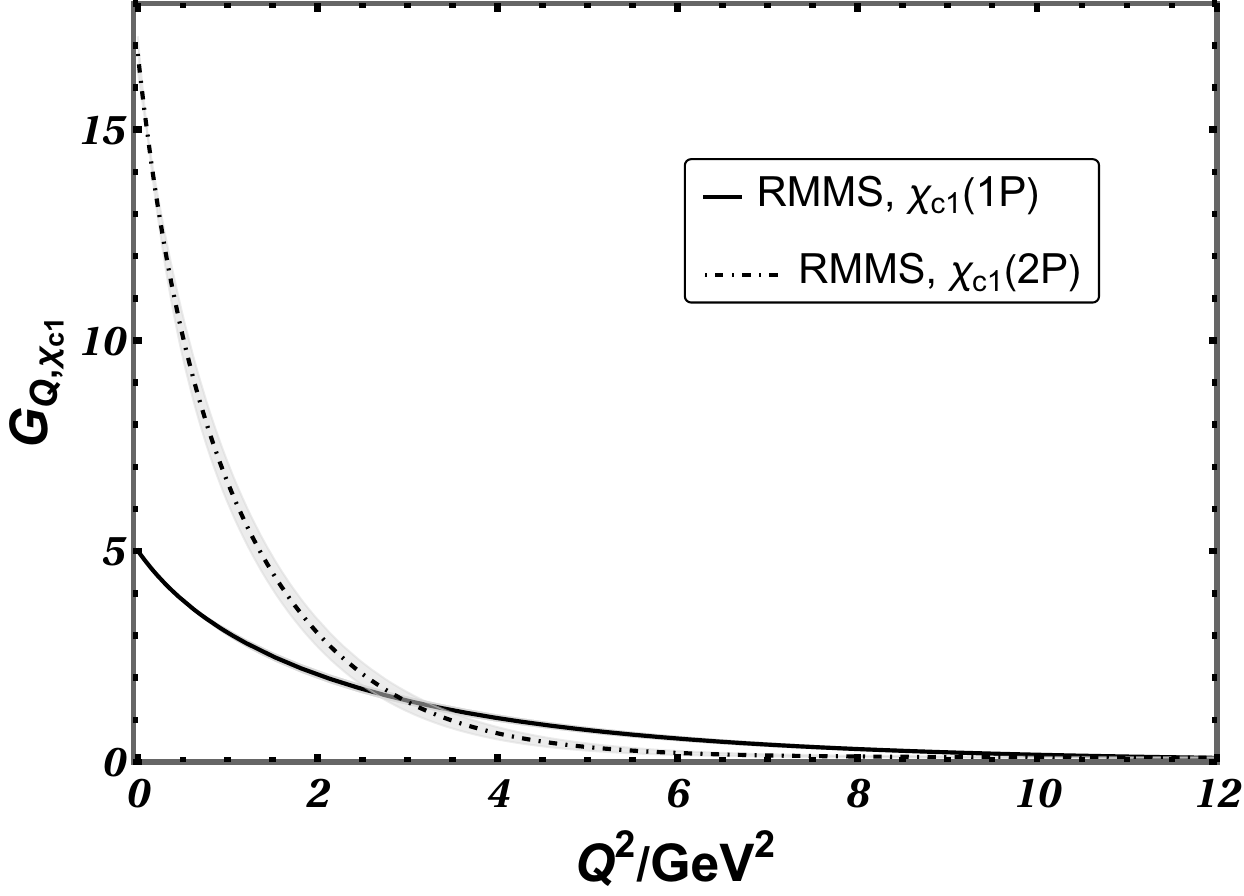}
 \caption{\label{fig:effchic11P2P} Single-quark electric charge (upper), magnetic dipole (middle) and electric quadrupole (lower) form factor of $\chi_{c1}(1P)$ (solid line) and $\chi_{c1}(2P)$ (dot-dashed line), and the shadow is the computational error. The dashed orange line (BLFQ2019) is the basis light front quantization result \cite{Adhikari2019}.}
\end{figure}

The single-quark electric form factors of $\chi_{c1}(1P)$ and $\chi_{c1}(2P)$ are displayed in the upper panel of Fig. \ref{fig:effchic11P2P}, the single-quark magnetic dipole form factors of $\chi_{c1}(1P)$ and $\chi_{c1}(2P)$ are displayed in the middle panel of Fig. \ref{fig:effchic11P2P}, the single-quark electric quadrupole form factors of $\chi_{c1}(1P)$ and $\chi_{c1}(2P)$ are displayed in the lower panel of Fig. \ref{fig:effchic11P2P}. Only one BLFQ result for $G_{C,\chi_{c1}(1P)}$ (BLFQ2019) is available for comparison; our result is lower than theirs. For the other results, no comparable data were found, and we provide our predictions.

Eq. (\ref{eq:gfactor}) shows that in the non-relativistic limit, the single-quark magnetic moment in unit of $\frac{e_q}{2m}$ in a $^3P_1$ meson is 1.5. Our results are $\mu_{1,\chi_{c1}(1P)} = G_{M,\chi_{c1}(1P)}(0)=1.45$, $\mu_{1,\chi_{c1}(2P)} = G_{M,\chi_{c1}(2P)}(0)=1.60$.

According to Eq. (\ref{eq:RMMS3P1}), $\chi_{c1}(1P)$ and $\chi_{c1}(2P)$ are predominantly $^3P_1$ states. Keeping contributions up to leading order and in the case of $\epsilon = \epsilon^+$, Eq. (\ref{eq:RMMS3P1}) is approximated as
\begin{eqnarray}\nonumber
 |A_1(p) \rangle_{\epsilon = \epsilon^+} &\approx &\sqrt{\frac{2E_{\bm{p}}}{N_c}} \sum\limits_{s\bar{s}}\int \frac{d^3\bm{k}}{(2\pi)^3}  \tilde{\varphi}_{A_1}\left(\bm{k}_r\right) b^\dag_{\bm{ks}} d^\dag_{\bm{\bar{k}\bar{s}}}| 0 \rangle,\\\label{eq:NRlimit3P1}
 \tilde{\varphi}_{A_1}\left(\bm{k}_r\right) &=& \varphi_{A_1}\left(|\bm{k}_r| \right)(\chi^{\bm{11}}_{\bm{s\bar{s}}} Y^{1,0} - \chi^{\bm{10}}_{\bm{s\bar{s}}} Y^{1,1})/\sqrt{2}
\end{eqnarray}
Eq. (\ref{eq:NRlimit3P1}) represents a prolate spheroidal-like charge distribution in coordinate space, hence the electric quadrupole moment is positive. Specifically, the charge distribution function is $\text{d}\rho(r,\theta) = \frac{3}{16\pi} \varphi^2_{A_1}\left(|\bm{r}| \right)(1+ \cos^2\theta) r^2\text{d} r\text{d}\Omega$, where $\theta$ is the azimuthal angle. According to our calculation, the single-quark electric quadrupole moments in unit of $\frac{e_q}{M^2}$ are $Q_{2,\chi_{c1}(1P)} = G_{Q,\chi_{c1}(1P)}(0) = 5.01$ and $Q_{2,\chi_{c1}(2P)} = G_{Q,\chi_{c1}(2P)}(0) = 16.8$.

\subsection{$h_c(1P)$ and $h_c(2P)$}\label{subsec:hc}

\begin{figure}[!t]
\centering
 \includegraphics[width=0.48\textwidth]{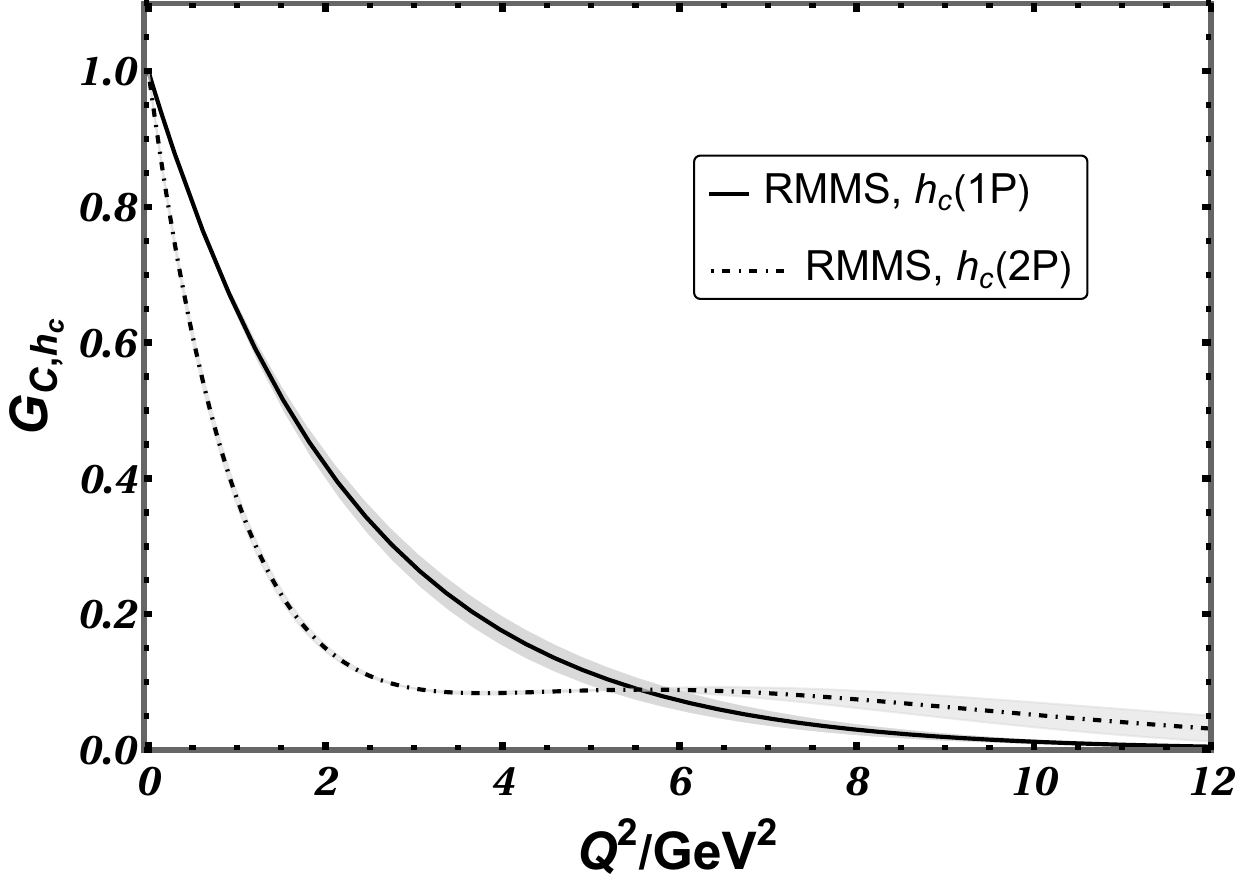}
 \includegraphics[width=0.48\textwidth]{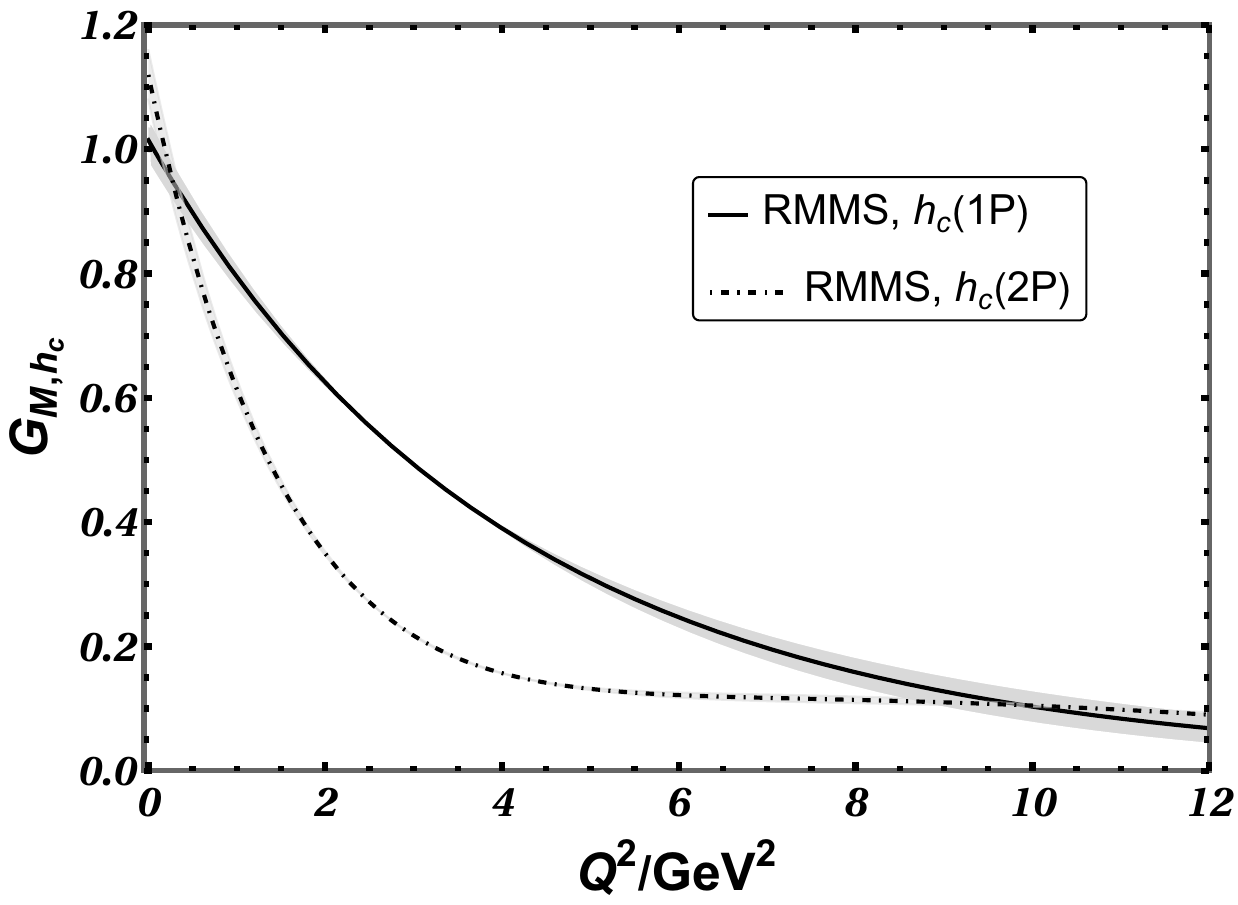}
 \includegraphics[width=0.48\textwidth]{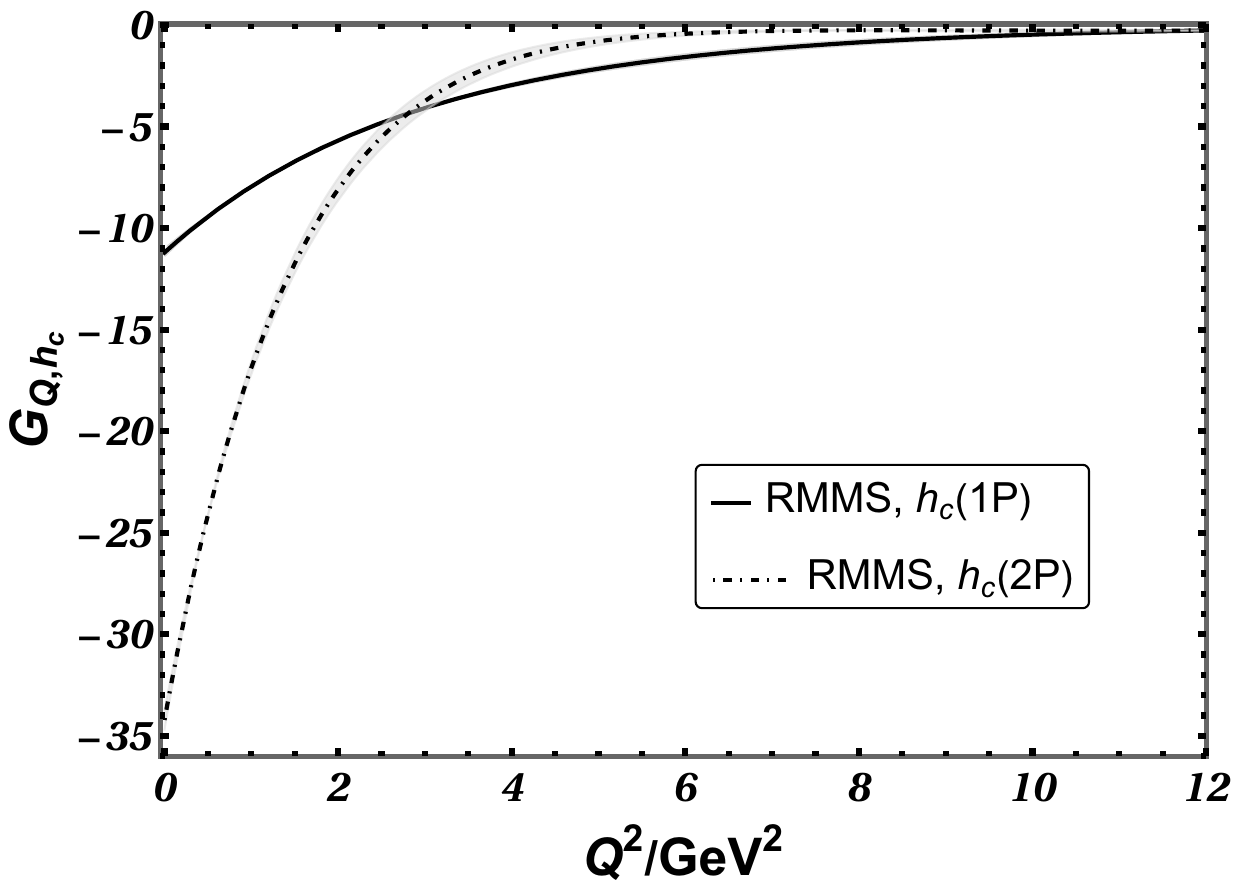}
 \caption{\label{fig:effhc1P2P} Single-quark electric charge (upper), magnetic dipole (middle) and electric quadrupole (lower) form factor of $h_{c}(1P)$ (solid line) and $h_{c}(2P)$ (dot-dashed line), and the shadow is the computational error.}
\end{figure}

The single-quark electric form factors of $h_c(1P)$ and $h_c(2P)$ are displayed in the upper panel of Fig. \ref{fig:effhc1P2P}, the single-quark magnetic dipole form factors of $h_c(1P)$ and $h_c(2P)$ are displayed in the middle panel of Fig. \ref{fig:effhc1P2P}, the single-quark electric quadrupole form factors of $h_c(1P)$ and $h_c(2P)$ are displayed in the lower panel of Fig. \ref{fig:effhc1P2P}. No comparable data were found, and we provide our predictions.

Eq. (\ref{eq:gfactor}) shows that in the non-relativistic limit, the single-quark magnetic moment in unit of $\frac{e_q}{2m}$ in a $^1P_1$ meson is 1.0. Our results are $\mu_{1,h_c(1P)} = G_{M,h_c(1P)}(0)=1.02$, $\mu_{1,h_c(2P)} = G_{M,h_c(2P)}(0)=1.12$.

According to Eq. (\ref{eq:RMMS1P1}), $h_c(1P)$ and $h_c(2P)$ are predominantly $^1P_1$ states. Keeping contributions up to leading order and in the case of $\epsilon = \epsilon^+$, Eq. (\ref{eq:RMMS1P1}) is approximated as
\begin{eqnarray}\nonumber
 |A_2(p) \rangle_{\epsilon = \epsilon^+} &\approx &\sqrt{\frac{2E_{\bm{p}}}{N_c}} \sum\limits_{s\bar{s}}\int \frac{d^3\bm{k}}{(2\pi)^3}  \tilde{\varphi}_{A_2}\left(\bm{k}_r\right) b^\dag_{\bm{ks}} d^\dag_{\bm{\bar{k}\bar{s}}}| 0 \rangle,\\\label{eq:NRlimit1P1}
 \tilde{\varphi}_{A_2}\left(\bm{k}_r\right) &=& \varphi_{A_2}\left(|\bm{k}_r| \right)\chi^{\bm{00}}_{\bm{s\bar{s}}} Y^{1,1}.
\end{eqnarray}
Eq. (\ref{eq:NRlimit1P1}) represents a oblate spheroidal-like charge distribution in coordinate space, hence the electric quadrupole moment is negative. Specifically, the charge distribution function is $\text{d}\rho(r,\theta) = \frac{3}{8\pi} \varphi^2_{A_2}\left(|\bm{r}| \right)\sin^2\theta r^2\text{d} r\text{d}\Omega$, where $\theta$ is the azimuthal angle. According to our calculation, the single-quark electric quadrupole moments in unit of $\frac{e_q}{M^2}$ are $Q_{2,h_c(1P)} = G_{Q,h_c(1P)}(0) = -11.2$ and $Q_{2,h_c(2P)} = G_{Q,h_c(2P)}(0) = -34.2$.

\subsection{$\chi_{c2}(1P)$ and $\chi_{c2}(2P)$}\label{subsec:chic2}

\begin{figure}[!t]
\centering
 \includegraphics[width=0.48\textwidth]{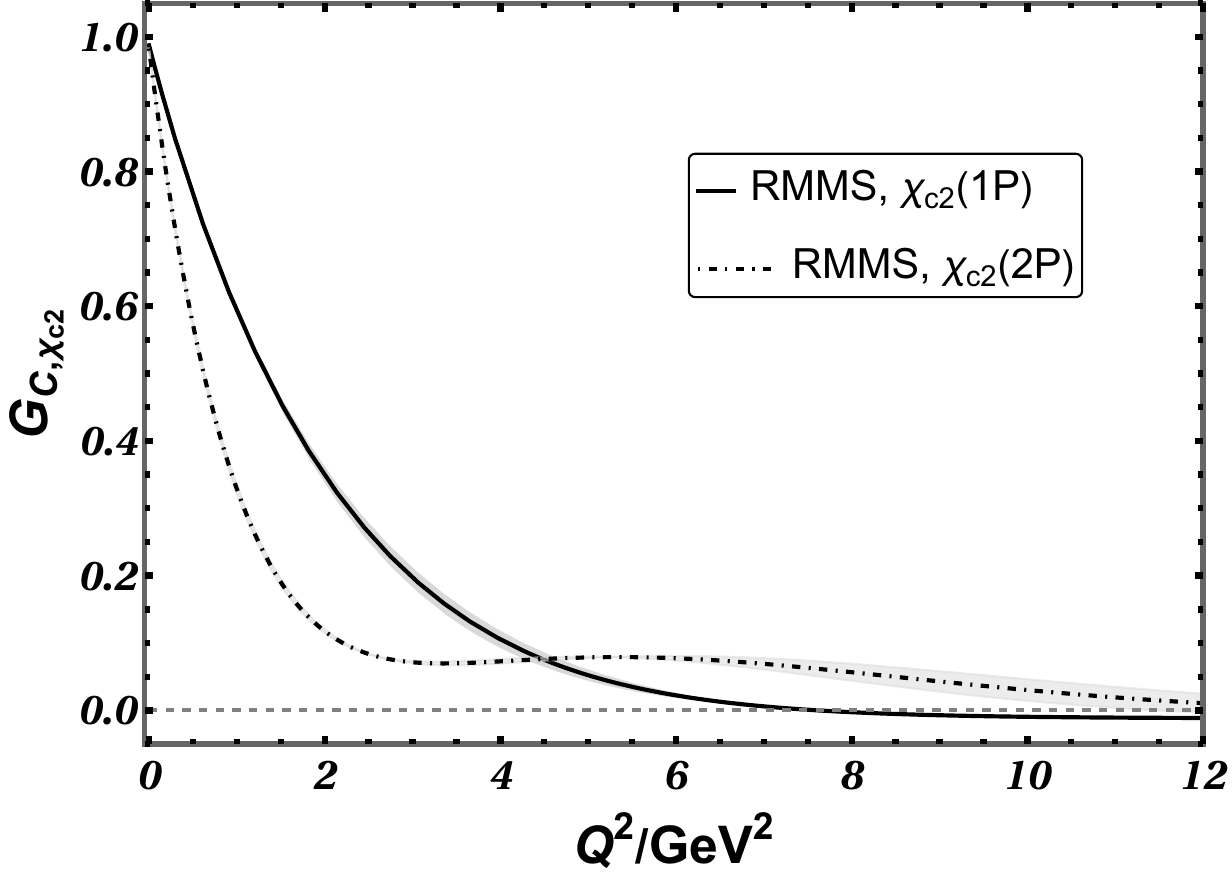}
 \includegraphics[width=0.48\textwidth]{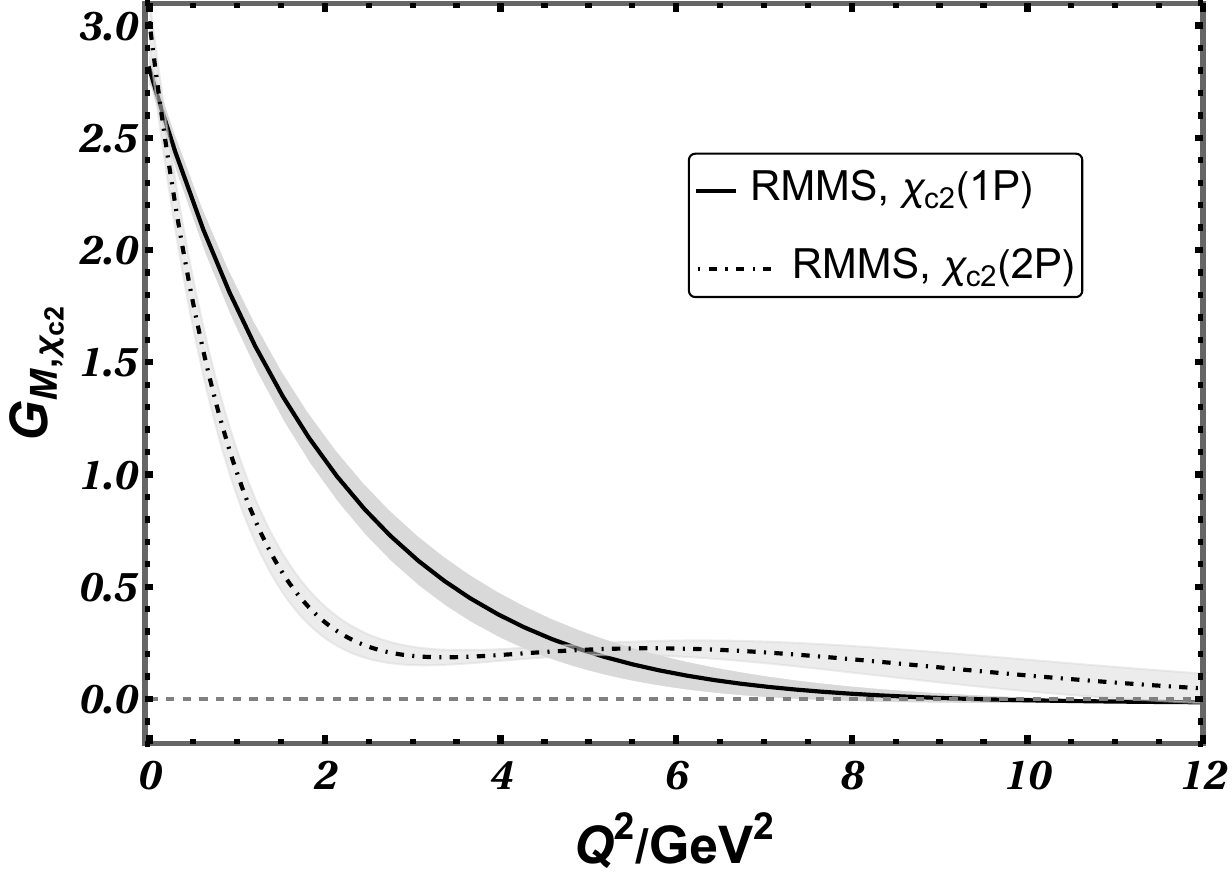}
 \includegraphics[width=0.48\textwidth]{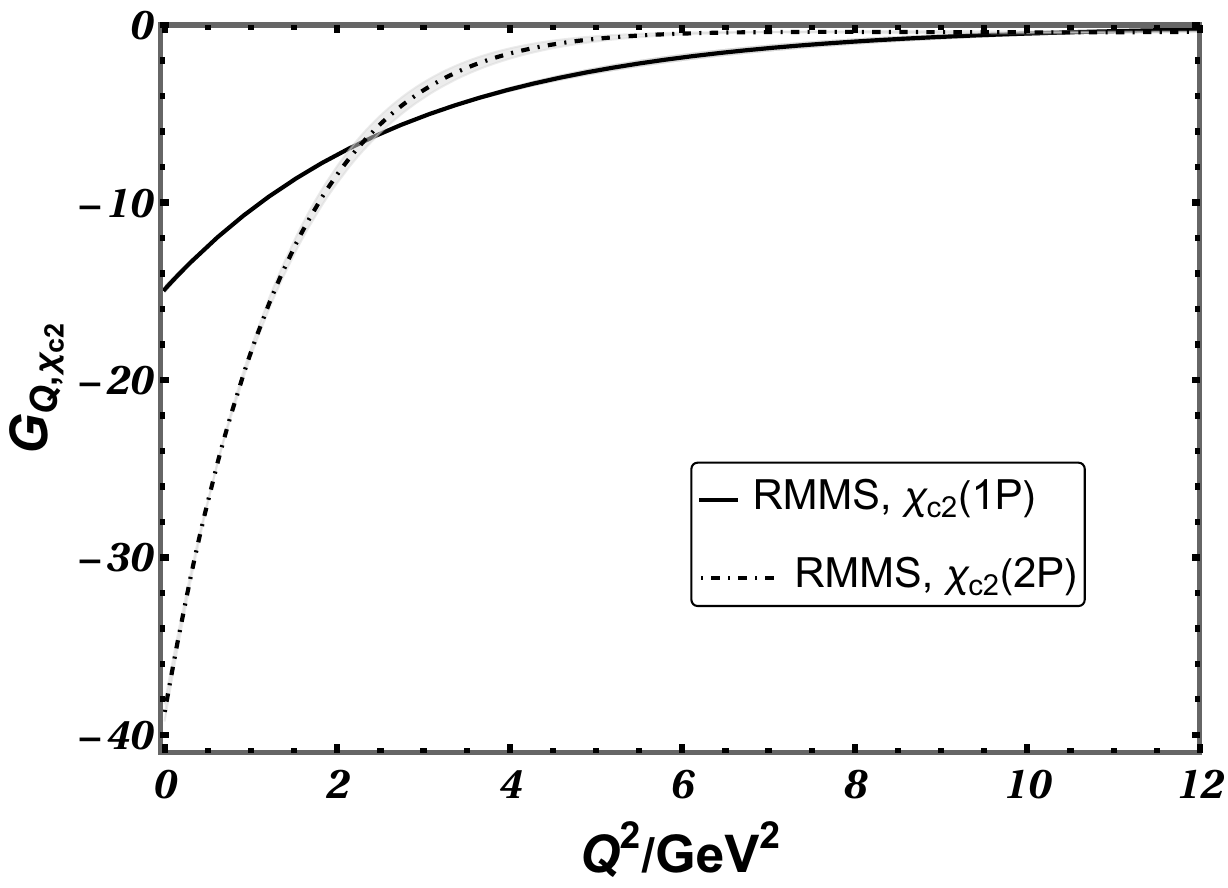}
 \caption{\label{fig:effchic21P2Pa} Single-quark electric charge (upper), magnetic dipole (middle) and electric quadrupole (lower) form factor of $\chi_{c2}(1P)$ (solid line) and $\chi_{c2}(2P)$ (dot-dashed line), and the shadow is the computational error.}
\end{figure}

\begin{figure}[!t]
\centering
 \includegraphics[width=0.48\textwidth]{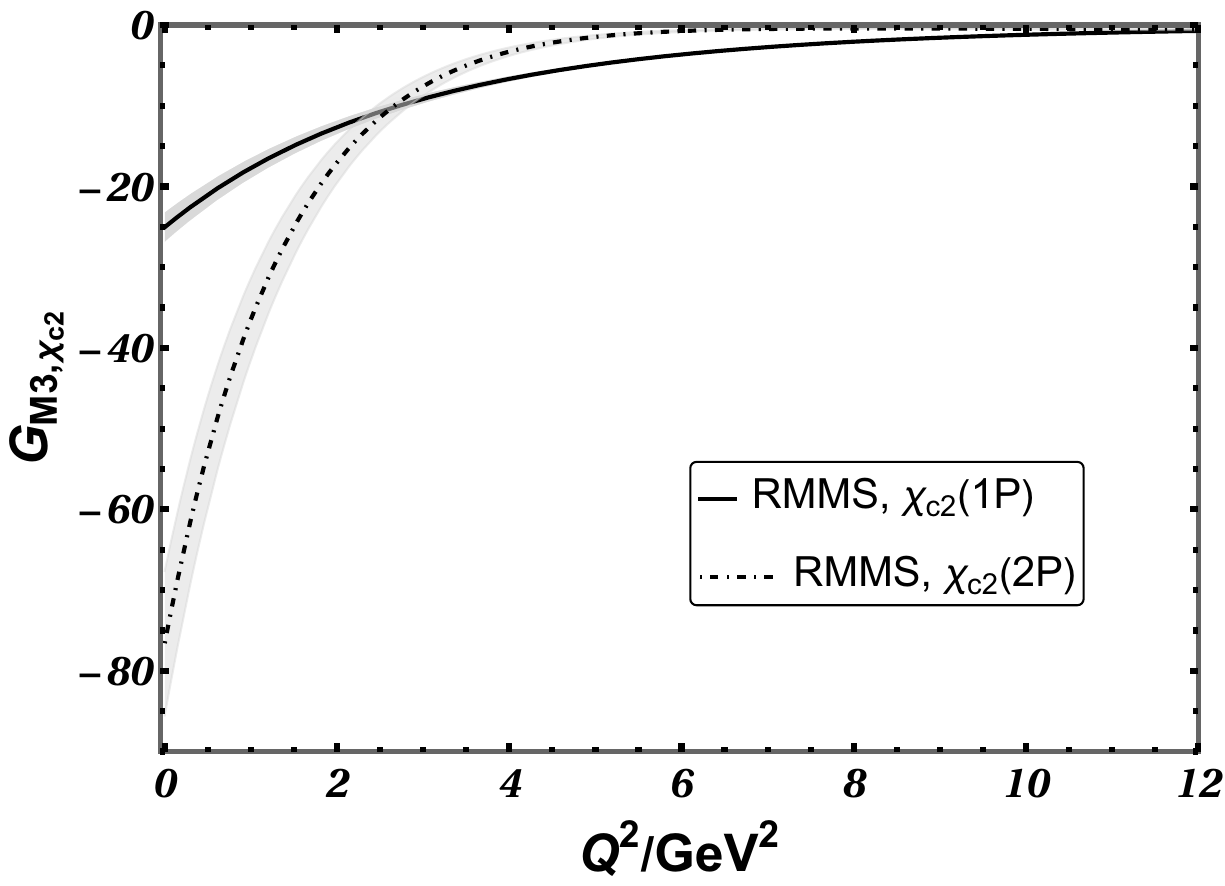}
 \includegraphics[width=0.48\textwidth]{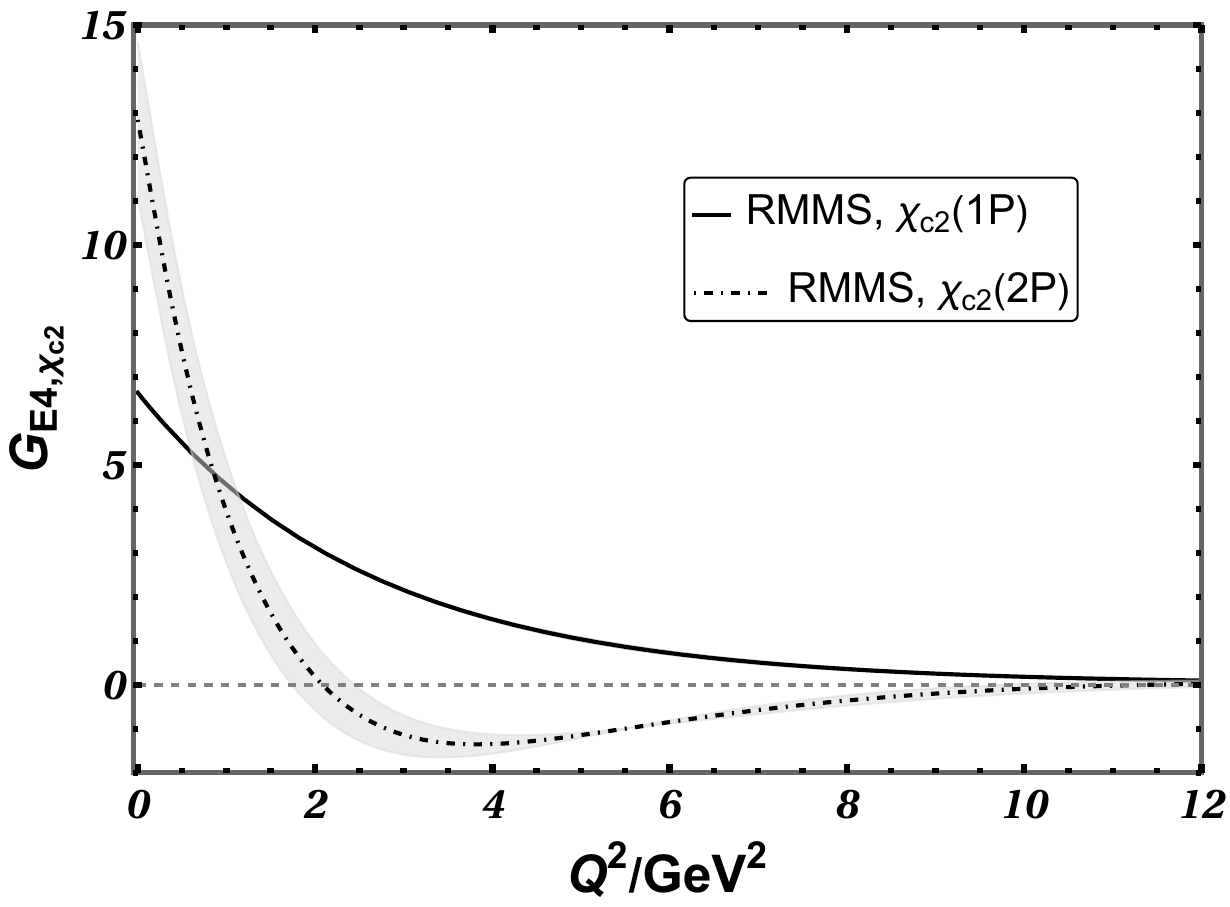}
 \caption{\label{fig:effchic21P2Pb} Single-quark magnetic octupole (upper) and electric hexadecapole form factor (lower) of $\chi_{c2}(1P)$ (solid line) and $\chi_{c2}(2P)$ (dot-dashed line), and the shadow is the computational error.}
\end{figure}

No comparable data were found for the single-quark electromagnetic form factors of $\chi_{c2}(1P)$ and $\chi_{c2}(2P)$, and we give our predictions. The single-quark electric form factors of $\chi_{c2}(1P)$ and $\chi_{c2}(2P)$ are displayed in the upper panel of Fig. \ref{fig:effchic21P2Pa}, the single-quark magnetic dipole form factors of $\chi_{c2}(1P)$ and $\chi_{c2}(2P)$ are displayed in the middle panel of Fig. \ref{fig:effchic21P2Pa}, the single-quark electric quadrupole form factors of $\chi_{c2}(1P)$ and $\chi_{c2}(2P)$ are displayed in the lower panel of Fig. \ref{fig:effchic21P2Pa}. The single-quark magnetic octupole form factors of $\chi_{c2}(1P)$ and $\chi_{c2}(2P)$ are displayed in the upper panel of Fig. \ref{fig:effchic21P2Pb}, the single-quark electric hexadecapole form factors of $\chi_{c2}(1P)$ and $\chi_{c2}(2P)$ are displayed in the lower panel of Fig. \ref{fig:effchic21P2Pb}.

Eq. (\ref{eq:gfactor}) shows that in the non-relativistic limit, the single-quark magnetic moment in unit of $\frac{e_q}{2m}$ in a $^3P_2$ meson is 3. Our results are $\mu_{1,\chi_{c2}(1P)} = G_{M,\chi_{c2}(1P)}(0)=2.82$, $\mu_{1,\chi_{c2}(2P)} = G_{M,\chi_{c2}(2P)}(0)=3.01$.

According to Eq. (\ref{eq:RMMS3P2}), $\chi_{c2}(1P)$ and $\chi_{c2}(2P)$ are predominantly $^3P_2$ states. Keeping contributions up to leading order and in the case of $\epsilon = \epsilon^{+2}$, Eq. (\ref{eq:RMMS3P2}) is approximated as
\begin{eqnarray}\nonumber
&& |T(p) \rangle_{\epsilon = \epsilon^{+2}} \approx \sqrt{\frac{2E_{\bm{p}}}{N_c}} \sum\limits_{s\bar{s}}\int \frac{d^3\bm{k}}{(2\pi)^3}  \tilde{\varphi}_{T}\left(\bm{k}_r\right) b^\dag_{\bm{ks}} d^\dag_{\bm{\bar{k}\bar{s}}}| 0 \rangle,\\\label{eq:NRlimit3P2}
&& \tilde{\varphi}_{T}\left(\bm{k}_r\right) = \varphi_{T}\left(|\bm{k}_r| \right)(\sqrt{\frac{2}{3}} \chi^{\bm{10}}_{\bm{s\bar{s}}} Y^{2,2}-\frac{1}{\sqrt{3}}\chi^{\bm{11}}_{\bm{s\bar{s}}} Y^{2,1})
\end{eqnarray}
Eq. (\ref{eq:NRlimit3P2}) represents a oblate spheroidal-like charge distribution in coordinate space, hence the electric quadrupole moment is negative. Specifically, the charge distribution function is $\text{d}\rho(r,\theta) = \frac{5}{16\pi} \varphi^2_{T}\left(|\bm{r}| \right)(1- \cos^4\theta) r^2\text{d} r\text{d}\Omega$, where $\theta$ is the azimuthal angle. According to our calculation, the single-quark electric quadrupole moments in unit of $\frac{e_q}{M^2}$ are $Q_{2,\chi_{c2}(1P)} = G_{Q,\chi_{c2}(1P)}(0) = -14.9$ and $Q_{2,\chi_{c2}(2P)} = G_{Q,\chi_{c2}(2P)}(0) = -38.7$.

Our calculation show that $\chi_{c2}(1P)$ and $\chi_{c2}(2P)$ possess a negative sigle-quark magnetic octupole moment and a positive sigle-quark electric hexadecapole moment. The magnetic octupole moment in unit of $\frac{e_q}{2M^3}$ is $\mu_{3,\chi_{c2}(1P)} = 9\times G_{M3,\chi_{c2}(1P)}(0) = -226$, $\mu_{3,\chi_{c2}(2P)} = 9\times G_{M3,\chi_{c2}(2P)}(0) = -689$. The electric hexadecapole moment in unit of $\frac{e_q}{M^4}$ is $Q_{4,\chi_{c2}(1P)} = 36\times G_{E4,\chi_{c2}(1P)}(0) = 239$, $Q_{4,\chi_{c2}(2P)} = 36\times G_{E4,\chi_{c2}(2P)}(0) = 461$.

\begin{table}[!t]
\caption{\label{tab:GM0GQ0} Summary of single-quark magnetic dipole moments and single-quark electric quadrupole moments. $n$ in the subscript is the radial quantum number, and $NR$ means non-relativistic limit.}
\begin{tabular}{c|c|c|c|c}
\hline
\makecell[c]{\vspace{1.0em}}\text{meson}& $J/\psi$ & $\chi_{c1}$ & $h_c$ & $\chi_{c2}$  \\
\hline
$\mu_{1,n=1}$ & 1.70 & 1.45& 1.02 & 2.82  \\
$\mu_{1,n=2}$ & 1.97 & 1.60& 1.12 & 3.01  \\
$\mu_{1,NR}$ & 2.0 & 1.5& 1.0 & 3.0  \\
$Q_{2,n=1}$ & -0.315 & 5.01& -11.2 & -14.9  \\
$Q_{2,n=2}$ & 0.033 & 16.8& -34.2 & -38.7  \\
\hline
\end{tabular}
\end{table}

Finally, for ease of comparison, we compile the results for the magnetic dipole moment and electric quadrupole moment in Table \ref{tab:GM0GQ0}.

\section{Summary and conclusion}\label{sec:conclusion}

In this work, we calculated the single-quark electromagnetic form factors of a broad subset of charmonium states, including $\eta_c(1S)$, $\eta_c(2S)$, $\chi_{c0}(1P)$, $\chi_{c0}(2P)$, $J/\psi(1S)$, $J/\psi(2S)$, $\chi_{c1}(1P)$, $\chi_{c1}(2P)$, $h_c(1P)$, $h_c(2P)$, $\chi_{c2}(1P)$ and $\chi_{c2}(2P)$, via a relativized quark model. We use the RMMS, Eq. (\ref{eq:RelativisedMockMeson}) - Eq. (\ref{eq:Relate3P2}), to present the mesons. The wave functions, Eq. (\ref{eq:Relate1S0}) - Eq. (\ref{eq:Relate3P2}), are determined by relating the non-relativistic limit of RMMS to the MMS, Eq. (\ref{eq:oldMockMeson}). Considering the reference frame dependence of our results, we take the result in the static frame  as the central value and treating the difference of the results between the Breit frame and the static frame as the computational error.

We compare our results with those of three groups of LQCD (Dudek2006 \cite{Dudek2006}, Chen2011 \cite{Chen2011} and Delaney2024 \cite{Delaney2024}), the Dyson-Schwinger equation (DSE2007 \cite{Maris2007}) and the basis light front quantization (BLFQ2019 \cite{Adhikari2019}) approaches whence available. Specifically,
\begin{itemize}
 \item our results and the DSE results are in good aggrement for $F_{\eta_c(1S)}$, $G_{C,J/\psi(1S)}$ and $G_{Q,J/\psi(1S)}$;
 \item our results and the LQCD results are in  reasonable aggrement for $F_{\eta_c(1S)}$, $F_{\eta_c(2S)}$, $F_{\chi_{c0}(1P)}$, $G_{C,J/\psi(1S)}$ and $G_{Q,J/\psi(1S)}$ considering the computational error.
 \item our results are lower than the BLFQ results for the radial ground state case, $F_{\eta_c(1S)}$, $F_{\chi_{c0}(1P)}$, $G_{C,J/\psi(1S)}$, $G_{M,J/\psi(1S)}$ and $G_{C,\chi_{c1}(1P)}$;
 \item our results and the BLFQ results are in good aggrement for the radial excited state case, $F_{\eta_c(2S)}$, $G_{C,J/\psi(2S)}$ and $G_{M,J/\psi(2S)}$;
 \item $G_{M,J/\psi(1S)}$ is the case where our result exhibits the greatest discrepancy with other results. Ref. \cite{Lakhina2006} also calculated $G_{M,J/\psi(1S)}$ via quark model using different parameters. It seems that this discrepancy cannot be reconciled by tuning parameters.
\end{itemize}
In summary, we calculate the single-quark electromagnetic form factors of the charmonium systematically. Our results are generally quite reasonable, considering that our calculated values of $F_{\eta_c(1S)}$, $F_{\eta_c(2S)}$, $F_{\chi_{c0}(1P)}$, $G_{C,J/\psi(1S)}$, $G_{Q,J/\psi(1S)}$, $G_{C,J/\psi(2S)}$ and $G_{M,J/\psi(2S)}$ are consistent with other theoretical results. We predict $F_{\chi_{c0}(2P)}$, $G_{Q,J/\psi(2S)}$, $G_{M,\chi_{c1}(1P)}$, $G_{Q,\chi_{c1}(1P)}$ and all the single-quark electromagnetic form factors of $\chi_{c1}(2P)$, $h_c(1P)$, $h_c(2P)$, $\chi_{c2}(1P)$ and $\chi_{c2}(2P)$ for the first time. The notable discrepancy between our $G_{M,J/\psi(1S)}$ result and other theoretical predictions requires further investigation.

\section*{Acknowledgments}
This work is supported by: the Science Foundation of education department of Hunan province, China under contracts No. 24B0067 and the National Natural Science Foundation of China (Grants No. 12175065 and 12235018).

\setcounter{section}{0}
\renewcommand{\thesection}{Appendix \Alph{section}}
\renewcommand{\thesubsection}{\Alph{section}\arabic{subsection}}

\section{Non-relativistic limit of the relativized mock meson state}\label{sec:appendixA}

\setcounter{equation}{0}
\renewcommand{\theequation}{A\arabic{equation}}
\setcounter{figure}{0}
\setcounter{table}{0}
\renewcommand{\thefigure}{A\arabic{figure}}
\renewcommand{\thetable}{A\arabic{table}}

The non-relativistic limits of the RMMS with $J^{PC} = 0^{-+}$ and $0^{++}$ have been given in the appendix of Ref. \cite{Ding2025}. In the following, we present the analogous derivations for $J^{PC} = 1^{--}$, $1^{++}$, $1^{+-}$ and $2^{++}$ states, respectively.

\subsection{$J^{PC} = 1^{--}$}

The vector RMMS is
\begin{eqnarray}\nonumber
|V(p) \rangle &=& \sqrt{\frac{2E_{\bm{p}}}{N_c}} \sum_{\bm{s,\bar{s}}} \int\frac{d^3\bm{k} d^3\bm{\bar{k}}}{(2\pi)^6}\frac{1}{\sqrt{2E_{\bm{k}}}} \frac{1}{\sqrt{2E_{\bar{\bm{k}}}}}\\\label{eq:RMMS3S1}
&&\hspace{-2em}\times\phi_V(k_r,p) \bar{u}(\bm{k},\bm{s}) \gamma^\mu v(\bar{\bm{k}},\bar{\bm{s}}) b^\dag_{\bm{ks}} d^\dag_{\bm{\bar{k}\bar{s}}} | 0 \rangle \cdot \epsilon_
\mu.
\end{eqnarray}
In the Dirac representation, the spinors are
\begin{eqnarray}
u(\bm{k},\bm{s}) &=& \sqrt{E_{\bm{k}} + m}
\begin{pmatrix}
 1 \\ \frac{\bm{\sigma}\cdot \bm{k}}{E_{\bm{k}} + m}
\end{pmatrix}
 \xi^s, \\
 v(\bar{\bm{k}},\bar{\bm{s}}) &=& \sqrt{\bar{E}_{\bar{\bm{k}}} + \bar{m}}
\begin{pmatrix}
\frac{\bm{\sigma}\cdot \bar{\bm{k}}}{\bar{E}_{\bar{\bm{k}}} + \bar{m}} \\ 1
\end{pmatrix}
 \eta^{\bar{s}},
\end{eqnarray}
where $\bm{\sigma}$ is the Pauli matrix, $\xi^s$ and $\eta^{\bar{s}}$ are two-compoment spinors for quark and antiquark. In the non-relativistic limit, $\bm{k} \to 0$ and $\bar{\bm{k}} \to 0$, Eq. (\ref{eq:RMMS3S1}) reduces to
\begin{eqnarray}\nonumber
|V(p) \rangle_{\epsilon=\epsilon^+} &\!\!=&\!\! \sqrt{\frac{2E_{\bm{p}}}{N_c}} \int\frac{d^3\bm{k} d^3\bm{\bar{k}}}{(2\pi)^6} \phi_V(k_r,p) \sqrt{2}\chi^{11}_{\bm{s\bar{s}}} b^\dag_{\bm{ks}} d^\dag_{\bm{\bar{k}\bar{s}}} | 0 \rangle,\\\nonumber
|V(p) \rangle_{\epsilon=\epsilon^0} &\!\!=&\!\! \sqrt{\frac{2E_{\bm{p}}}{N_c}} \int\frac{d^3\bm{k} d^3\bm{\bar{k}}}{(2\pi)^6} \phi_V(k_r,p) \sqrt{2}\chi^{10}_{\bm{s\bar{s}}} b^\dag_{\bm{ks}} d^\dag_{\bm{\bar{k}\bar{s}}} | 0 \rangle,\\\nonumber
|V(p) \rangle_{\epsilon=\epsilon^-} &\!\!=&\!\! \sqrt{\frac{2E_{\bm{p}}}{N_c}} \int\frac{d^3\bm{k} d^3\bm{\bar{k}}}{(2\pi)^6} \phi_V(k_r,p) \sqrt{2}\chi^{1,-1}_{\bm{s\bar{s}}} b^\dag_{\bm{ks}} d^\dag_{\bm{\bar{k}\bar{s}}} | 0 \rangle,\\\label{eq:ReducedRMMS3S1}
\end{eqnarray}
where $\epsilon^{\pm} = (0, \mp\frac{1}{\sqrt{2}}, -\frac{i}{\sqrt{2}}, 0)$ and $\epsilon^{0} = (0,0,0,1)$ are the polarization vectors, $\chi^{11}_{\bm{s\bar{s}}} = \xi^\uparrow\eta^\uparrow $, $\chi^{10}_{\bm{s\bar{s}}} = \frac{1}{\sqrt{2}} (\xi^\uparrow\eta^\downarrow + \xi^\downarrow\eta^\uparrow)$ and $\chi^{11}_{\bm{s\bar{s}}} = \xi^\downarrow\eta^\downarrow $ are the triplet spin wave functions. Comparing Eq. (\ref{eq:ReducedRMMS3S1}) with Eq. (\ref{eq:oldMockMeson}), we get
\begin{equation}\label{eq:relate3S1prime}
 \phi_V(k_r,p)= \frac{\varphi_V\left(|\bm{k}_r|\right)}{\sqrt{8\pi}} (2\pi)^3\delta^{(3)}(\bm{k}+\bm{\bar{k}}-\bm{p}).
\end{equation}
In order to ensure the normalization condition of electric charge form factor, $G_{E}(0)=1$, a factor $\sqrt{\frac{2E_{\bm{k}} \bar{E}_{\bar{\bm{k}}}}{E_{\bm{k}}\bar{E}_{\bar{\bm{k}}} +m\bar{m} +\bm{k}^2_r/3}}$ is multiply to the left of Eq. (\ref{eq:relate3S1prime}).

\subsection{$J^{PC} = 1^{++}$}

The RMMS with $J^{PC} = 1^{++}$ is
\begin{eqnarray}\nonumber
|A_1(p) \rangle &=& \sqrt{\frac{2E_{\bm{p}}}{N_c}} \sum_{\bm{s,\bar{s}}} \int\frac{d^3\bm{k} d^3\bm{\bar{k}}}{(2\pi)^6}\frac{1}{\sqrt{2E_{\bm{k}}}} \frac{1}{\sqrt{2E_{\bar{\bm{k}}}}}\\\label{eq:RMMS3P1}
&&\hspace{-4em}\times\phi_{A1}(k_r,p) \bar{u}(\bm{k},\bm{s})\gamma^\mu\gamma_5 v(\bar{\bm{k}},\bar{\bm{s}}) b^\dag_{\bm{ks}} d^\dag_{\bm{\bar{k}\bar{s}}} | 0 \rangle\cdot \epsilon_
\mu.
\end{eqnarray}
In the non-relativistic limit, Eq. (\ref{eq:RMMS3P1}) reduces to
\begin{eqnarray}\nonumber
|A_1(p) \rangle &=& \sqrt{\frac{2E_{\bm{p}}}{N_c}} \sum_{\bm{s,\bar{s}}} \int\frac{d^3\bm{k} d^3\bm{\bar{k}}}{(2\pi)^6} \sqrt{\frac{8\pi}{3m\bar{m}}}|\bm{k}_r|\\\label{eq:ReducedRMMS3P1}
&&\hspace{-2em}\times\phi_{A_1}(k_r,p)  \langle\chi^{1m_S}_{\bm{s\bar{s}}}Y^{1m_L};1m_J\rangle b^\dag_{\bm{ks}} d^\dag_{\bm{\bar{k}\bar{s}}} | 0 \rangle,
\end{eqnarray}
where $\langle\chi^{1m_S}_{\bm{s\bar{s}}}Y^{1m_L};1m_J\rangle$ means the triplet spin wave function $\chi^{1m_S}$ and the first-order spherical harmonic function $Y^{1m_L}$ being combined by the Clebsch-Gordan coefficient to form a $J=1$ state, with the magnetic quantum number $m_J$ determined by the polarization vector. Comparing Eq. (\ref{eq:ReducedRMMS3P1}) with Eq. (\ref{eq:oldMockMeson}), we get
\begin{equation}\label{eq:relate3P1prime}
 \phi_{A1}(k_r,p)= \frac{\sqrt{3m\bar{m}}\varphi_{A_1}\left(|\bm{k}_r|\right)}{\sqrt{8\pi}|\bm{k}_r|} (2\pi)^3\delta^{(3)}(\bm{k}+\bm{\bar{k}}-\bm{p}).
\end{equation}
In order to ensure the normalization condition of electric charge form factor, $G_{E}(0)=1$, a factor $\sqrt{\frac{4E_{\bm{k}} \bar{E}_{\bar{\bm{k}}} \bm{k}^2_r }{ 3m\bar{m} (E_{\bm{k}}\bar{E}_{\bar{\bm{k}}} -m\bar{m} +\bm{k}^2_r/3)}}$ is multiply to the left of Eq. (\ref{eq:relate3P1prime}).

\subsection{$J^{PC} = 1^{+-}$}

The RMMS with $J^{PC} = 1^{+-}$ is
\begin{eqnarray}\nonumber
|A_2(p) \rangle &=& \sqrt{\frac{2E_{\bm{p}}}{N_c}} \sum_{\bm{s,\bar{s}}} \int\frac{d^3\bm{k} d^3\bm{\bar{k}}}{(2\pi)^6}\frac{1}{\sqrt{2E_{\bm{k}}}} \frac{1}{\sqrt{2E_{\bar{\bm{k}}}}}\\\label{eq:RMMS1P1}
&&\hspace{-4em}\times\phi_{A2}(k_r,p) \bar{u}(\bm{k},\bm{s})k_r^\mu\gamma_5 v(\bar{\bm{k}},\bar{\bm{s}}) b^\dag_{\bm{ks}} d^\dag_{\bm{\bar{k}\bar{s}}} | 0 \rangle\cdot \epsilon_
\mu.
\end{eqnarray}
In the non-relativistic limit, Eq. (\ref{eq:RMMS1P1}) reduces to
\begin{eqnarray}\nonumber
|A_2(p) \rangle &=& \sqrt{\frac{2E_{\bm{p}}}{N_c}} \sum_{\bm{s,\bar{s}}} \int\frac{d^3\bm{k} d^3\bm{\bar{k}}}{(2\pi)^6} \sqrt{\frac{8\pi}{3}}|\bm{k}_r|\\\label{eq:ReducedRMMS1P1}
&&\times\phi_{A_2}(k_r,p) Y^{1m_J} b^\dag_{\bm{ks}} d^\dag_{\bm{\bar{k}\bar{s}}} | 0 \rangle,
\end{eqnarray}
where the magnetic quantum number $m_J$ is determined by the polarization vector. Comparing Eq. (\ref{eq:ReducedRMMS1P1}) with Eq. (\ref{eq:oldMockMeson}), we get
\begin{equation}\label{eq:relate1P1prime}
 \phi_{A2}(k_r,p)= \frac{\sqrt{3}\varphi_{A_2}\left(|\bm{k}_r|\right)}{\sqrt{8\pi}|\bm{k}_r|} (2\pi)^3\delta^{(3)}(\bm{k}+\bm{\bar{k}}-\bm{p}).
\end{equation}
Eq. (\ref{eq:relate3P1prime}) already satisties the normalization condition of electric charge form factor, $G_{E}(0)=1$.

\subsection{$J^{PC} = 2^{++}$}

The RMMS with $J^{PC} = 2^{++}$ is
\begin{eqnarray}\nonumber
|T(p) \rangle &=& \sqrt{\frac{2E_{\bm{p}}}{N_c}} \sum_{\bm{s,\bar{s}}} \int\frac{d^3\bm{k} d^3\bm{\bar{k}}}{(2\pi)^6}\frac{1}{\sqrt{2E_{\bm{k}}}} \frac{1}{\sqrt{2E_{\bar{\bm{k}}}}}\\\label{eq:RMMS3P2}
&&\hspace{-6em}\times\phi_{T}(k_r,p) \bar{u}(\bm{k},\bm{s})(\gamma^\mu k_r^\nu\!\! +\!\! \gamma^\nu k_r^\mu) v(\bar{\bm{k}},\bar{\bm{s}}) b^\dag_{\bm{ks}} d^\dag_{\bm{\bar{k}\bar{s}}} | 0 \rangle\!\!\cdot\!\! \epsilon_
\mu.
\end{eqnarray}
In the non-relativistic limit, Eq. (\ref{eq:RMMS3P2}) reduces to
\begin{eqnarray}\nonumber
|T(p) \rangle &=& \sqrt{\frac{2E_{\bm{p}}}{N_c}} \sum_{\bm{s,\bar{s}}} \int\frac{d^3\bm{k} d^3\bm{\bar{k}}}{(2\pi)^6} \sqrt{\frac{32\pi}{3}}|\bm{k}_r|\\\label{eq:ReducedRMMS3P2}
&&\hspace{-2em}\times\phi_{T}(k_r,p)  \langle\chi^{1m_S}_{\bm{s\bar{s}}}Y^{1m_L};2m_J\rangle b^\dag_{\bm{ks}} d^\dag_{\bm{\bar{k}\bar{s}}} | 0 \rangle,
\end{eqnarray}
where $\langle\chi^{1m_S}_{\bm{s\bar{s}}}Y^{1m_L};2m_J\rangle$ means the triplet spin wave function $\chi^{1m_S}$ and the first-order spherical harmonic function $Y^{1m_L}$ being combined by the Clebsch-Gordan coefficient to form a $J=2$ state, with the magnetic quantum number $m_J$ determined by the polarization tensor. Comparing Eq. (\ref{eq:ReducedRMMS3P2}) with Eq. (\ref{eq:oldMockMeson}), we get
\begin{equation}\label{eq:relate3P2prime}
 \phi_{T}(k_r,p)= \frac{\sqrt{3}\varphi_{T}\left(|\bm{k}_r|\right)}{\sqrt{32\pi}|\bm{k}_r|} (2\pi)^3\delta^{(3)}(\bm{k}+\bm{\bar{k}}-\bm{p}).
\end{equation}
In order to ensure the normalization condition of electric charge form factor, $G_{E}(0)=1$, a factor $\sqrt{\frac{2E_{\bm{k}} \bar{E}_{\bar{\bm{k}}}}{E_{\bm{k}}\bar{E}_{\bar{\bm{k}}} +m\bar{m} +\bm{k}^2_r/3}}$ is multiply to the left of Eq. (\ref{eq:relate3P2prime}).

\section{The procedure for calculating the single-quark form factors}\label{sec:appendixB}

\setcounter{equation}{0}
\renewcommand{\theequation}{B\arabic{equation}}
\setcounter{figure}{0}
\setcounter{table}{0}
\renewcommand{\thefigure}{B\arabic{figure}}
\renewcommand{\thetable}{B\arabic{table}}

The expressions of the electromagnetic form factors using the MMS, Eq. (\ref{eq:oldMockMeson}), have been given in Ref. \cite{Lakhina2006}. In this appendix we explain the calculation procedure for the form factors using RMMS.

\subsection{$J=0$ meson}

In the case of pseudoscalar and scalar meson, $J=0$, so there is only one form factor, $F_M(Q^2)$. It can be obtained by computing the matrix element, $\langle M(P_2)| j^\mu(0)|M(P_1)\rangle$. By utilizing the commutation relations of creation and annihilation operators and the operational properties of Dirac spinors, we get the matrix element for a single quark
\begin{eqnarray}\nonumber
& \langle M(P_2)| j^\mu(0)|M(P_1)\rangle_{\text{sq}} =\frac{\sqrt{E_{\bm{P}_1}E_{\bm{P}_2}}}{N_c} \int\frac{d^3 \bm{k}_1}{(2\pi)^3} \frac{d^3 \bar{\bm{k}}_1}{(2\pi)^3} \frac{d^3 \bm{k}_2}{(2\pi)^3}\\\nonumber
& \text{tr}[(\slashed{\bar{k}}_1 - \bar{m})\Gamma_M(k_{2r},P_2) (\slashed{k}_2 + m)\gamma^\mu (\slashed{k}_1 + m)\Gamma_M(k_{1r},P_1)],\\
\end{eqnarray}
where tr means the trace over the Dirac matrices,他 $\bm{k}_1$ and $\bar{\bm{k}}_1$ are the 3-momenta of the quark and antiquark in the initial meson, $\bm{k}_2$ is the 3-momenta of the quark in the initial meson, $k_1 = (E_{\bm{k}_1},\bm{k}_1)$, $\bar{k}_1 = (\bar{E}_{\bar{\bm{k}}_1},\bar{\bm{k}}_1)$, $k_2 = (E_{\bm{k}_2},\bm{k}_2)$, $k_{1r} = \frac{\bar{m}k_1 - m\bar{k}_1}{m+\bar{m}}$ and $k_{2r} = \frac{\bar{m}k_2 - m\bar{k}_2}{m+\bar{m}}$ are the relative 4-momenta, with $\bar{\bm{k}}_2 = \bar{\bm{k}}_1$.

\subsection{$J=1$ meson}

Assuming the meson moves in the z-direction,
\begin{equation}
 P_1 = (E_{\bm{P}_1}, 0, 0, |\bm{P}_1|),\quad P_2 = (E_{\bm{P}_2}, 0, 0, |\bm{P}_2|),
\end{equation}
the polarization vectors of $J=1$ meson are
\begin{eqnarray}\nonumber
 \epsilon_1^\pm = \frac{1}{\sqrt{2}}(0,\pm 1, -i,0), & \epsilon_1^0 = \frac{1}{M}(|\bm{P}_1|, 0, 0, E_{\bm{P}_1}),\\
  \epsilon_2^\pm = \frac{1}{\sqrt{2}}(0,\pm 1, -i,0), & \epsilon_2^0 = \frac{1}{M}(|\bm{P}_2|, 0, 0, E_{\bm{P}_2}).
\end{eqnarray}
In order to calculate the electromagnetic form factors of $J=1$ meson, we need three nonzero and independent matrix elements. There exist three and only three,
\begin{equation}
 \langle \epsilon_2^+\epsilon_1^+\rangle = \langle \epsilon_2^-\epsilon_1^-\rangle,\;
 \langle \epsilon_2^\pm\epsilon_1^0\rangle = \langle \epsilon_2^0\epsilon_1^\pm\rangle,\;
  \langle \epsilon_2^0\epsilon_1^0\rangle ,
\end{equation}
where $\langle \epsilon_2\epsilon_1\rangle $ is short for $ \langle M(P_2,\epsilon_2)| j^\mu(0)|M(P_1,\epsilon_1)\rangle$.

\subsection{$J=2$ meson}

The polarization tensors of $J=2$ meson are
\begin{eqnarray}\nonumber
 \epsilon^{\pm 2}_{\mu\nu} &=& \epsilon^{\pm}_{\mu} \epsilon^{\pm}_{\nu}\\
 \epsilon^{\pm 1}_{\mu\nu} &=& \frac{1}{\sqrt{2}} (\epsilon^{\pm}_{\mu} \epsilon^{0}_{\nu} + \epsilon^{0}_{\mu} \epsilon^{\pm}_{\nu} )\\\nonumber
 \epsilon^{0}_{\mu\nu} &=& \frac{1}{\sqrt{6}} ( \epsilon^{+}_{\mu} \epsilon^{-}_{\nu} + 2\epsilon^{0}_{\mu} \epsilon^{0}_{\nu} + \epsilon^{-}_{\mu} \epsilon^{+}_{\nu} ).
\end{eqnarray}
In order to calculate the electromagnetic form factors of $J=2$ meson, we need five nonzero and independent matrix elements. There exist five and only five,
\begin{eqnarray}\nonumber
&& \langle \epsilon_2^{+2}\epsilon_1^{+2}\rangle = \langle \epsilon_2^{-2}\epsilon_1^{-2}\rangle,\\\nonumber
&& \langle \epsilon_2^{+2}\epsilon_1^{+1}\rangle =\langle \epsilon_2^{+1}\epsilon_1^{+2}\rangle = \langle \epsilon_2^{-2}\epsilon_1^{-1}\rangle= \langle \epsilon_2^{-1}\epsilon_1^{-2}\rangle,\\
&& \langle \epsilon_2^{+1}\epsilon_1^{+1}\rangle = \langle \epsilon_2^{-1}\epsilon_1^{-1}\rangle,\\\nonumber
&& \langle \epsilon_2^{+1}\epsilon_1^{0}\rangle =\langle \epsilon_2^{0}\epsilon_1^{+1}\rangle = \langle \epsilon_2^{-1}\epsilon_1^{0}\rangle= \langle \epsilon_2^{0}\epsilon_1^{-1}\rangle,\\\nonumber
&& \langle \epsilon_2^{0}\epsilon_1^{0}\rangle.
\end{eqnarray}

\bibliographystyle{unsrt}
\bibliography{../reference/EFFcc}

\end{document}